\documentclass[11pt]{article}
\usepackage[utf8]{inputenc}
\usepackage{setspace}

\usepackage[margin=1in]{geometry}

\usepackage{apacite}
\usepackage{blindtext}
\usepackage{amsmath,bm}
\usepackage{color}
\usepackage{rotating}

\usepackage{caption}
\usepackage{subcaption}

\usepackage{grffile}

\usepackage{graphicx}

\usepackage{lscape}

\usepackage{blindtext}

\usepackage{amssymb}
\usepackage{varioref} 
\usepackage{textcomp}

\usepackage{booktabs}

\usepackage{multirow}

\usepackage{pdflscape}

\usepackage{relsize}
\usepackage{tablefootnote}
\usepackage{wrapfig}
\usepackage{natbib}
\usepackage{listings}
\usepackage{longtable}
\usepackage{url}
\usepackage{tabularx}
\usepackage{float}
\usepackage{epsfig}
\usepackage[titletoc]{appendix}
\usepackage{arydshln}
\usepackage{footnote}
\usepackage{secdot}
\usepackage{bigints}
\usepackage[utf8]{inputenc}
\usepackage[english]{babel}
\usepackage{xr}
\usepackage{adjustbox}
\usepackage{graphicx}

\usepackage{multicol}
\usepackage{enumitem}

\usepackage{bigstrut}
\usepackage{rotating}
\usepackage{xr}

\title{A Support Vector Machine Based Cure Rate Model For Interval Censored Data}
\author{S. Pal\thanks{S. Pal is with Department of Mathematics, University of Texas at Arlington,  Texas, USA (email: suvra.pal@uta.edu).}, Y. Peng\thanks{Y. Peng is with Department of Public Health Sciences, Queen’s University, Kingston, Ontario, Canada}, S. Barui\thanks{S. Barui is with Quantitative Methods and Operations Management Area, Indian Institute of Management, Kozhikode, Kerala, India. } and P. Wang\thanks{P. Wang is with Department of Mathematics, University of Texas at Arlington, Texas, USA}}
\date{}

\begin{document}
\maketitle
\begin{abstract}
\noindent The mixture cure rate model is the most commonly used cure rate model in the literature. In the context of mixture cure rate model, the standard approach to model the effect of covariates on the cured or uncured probability is to use a logistic function. This readily implies that the boundary classifying the cured and uncured subjects is linear. In this paper, we propose a new mixture cure rate model based on interval censored data that uses the support vector machine (SVM) to model the effect of covariates on the uncured or the cured probability (i.e., on the incidence part of the model). Our proposed model inherits the features of the SVM and provides flexibility to capture classification boundaries that are non-linear and more complex. Furthermore, the new model can be used to model the effect of covariates on the incidence part when the dimension of covariates is high. The latency part is modeled by a   proportional hazards structure. We develop an estimation procedure based on the expectation maximization (EM) algorithm to estimate the cured/uncured probability and the latency model parameters. Our simulation study results show that the proposed model performs better in capturing complex classification boundaries when compared to the existing logistic regression based  mixture cure rate model. We also show that our model's ability to capture complex classification boundaries improve the estimation results corresponding to the latency parameters. For illustrative purpose, we present our analysis by applying the proposed methodology to an interval censored data on smoking cessation.
\end{abstract}

\textbf{Keywords}: Support vector machine; Multiple imputation; Sequential minimal optimization; Mixture cure rate model; EM algorithm

\section{Introduction}\label{sect1}   
Ordinary survival analysis techniques such as the proportional hazards (PH) model, the proportional odds (PO) model or the accelerated failure time (AFT) model are concerned with modeling censored time-to-event data by assuming that every subject in the study will encounter the primary event of interest (death, relapse, or recurrence of a disease etc.). However, it is not appropriate to apply these techniques to situations where a portion of the study cohort does not experience the event, e.g., clinical studies involving low fatality rate with death as the event. It can be argued that if these subjects are followed up sufficiently beyond the study period, they may face the event due to some other risk factors. Therefore, these subjects can be considered as cured with respect to the event of interest. The survival model that incorporates the effects of such cured subjects is called the cure rate model. Remarkable progress in medical sciences also necessitate further exploration in to the cure rate model where estimating the cure fraction precisely can be of great importance \cite{PengYu21}. \\

Introduced by \citet{boag1949maximum} and exclusively studied by \citet{berkson1952survival}, the mixture cure rate model is perhaps the most popular cure rate model. If $T^*$ denotes the lifetime of a susceptible (not cured) subject, then, the actual lifetime $T$ for any subject can be modeled by 
\begin{equation}\label{eq1}
T=JT^*+(1-J)\infty,
\end{equation} 
where $J$ is a cure indicator denoting if an individual is cured ($J=0$) or not ($J=1$). Further, considering $S_p(t)=P(T>t)$ and $S_u(t)=P(T^*>t)$ as the respective survival functions corresponding to $T$ and $T^*$, we can express
\begin{equation}\label{eq2}
S_p(t)=(1-\pi) + \pi S_u(t),
\end{equation} 
where $\pi=P(J=1)$. The latency part $S_u(t)=S_u(t|\bm x)$ and the incidence part $\pi=\pi(\bm z)$ are generally modeled to incorporate the effects of covriates $\bm x=(x_1, \dots, x_p)^{\tiny \rm T}$ and $\bm z=(z_1, \dots, z_q)^{\tiny \rm T}$ for any integers $p$ and $q$. Note here that $\bm x$ and $\bm z$ may share the same covariates.  \\

The properties of the mixture cure rate model  with various assumptions and extensions are explored in details by several authors. Modeling lifetime of the susceptible individuals have been studied extensively. For example, a complete parametric mixture cure rate model is  studied by \cite{farewell1982use, farewell1986mixture} by assuming homogeneous Weibull lifetimes and logit-link to the cure rate. Semiparametric cure models with PH structure of the latency is studied extensively by \cite{kuk1992mixture}, \cite{peng2000nonparametric} and  \cite{sy2000estimation}, to name a few. Generalizations to semiparametric PO (\citealp{gu2011analysis, mao2010semiparametric}), AFT (\citealp{li2002semi, zhang2007new, zhang2009accelerated}), transformation class (\citealp{lu2004semiparametric}) and additive hazards (\citealp{barui2020semiparametric}) under mixture cure rate model were also investigated with various estimation techniques and model considerations. \\

On the other hand, the incidence part $\pi (\bm z)$ is traditionally and extensively modeled by sigmoid or logistic function
\begin{equation}\label{eq3}
\pi(\bm z)=\frac{\exp({\bm z^{*\tiny \rm T} \bm \beta})}{1+\exp({\bm z^{*\tiny \rm T} \bm \beta})},
\end{equation} 
where $\bm \beta=(\beta_0, \beta_1, \dots, \beta_q)^{\tiny \rm T}$ and $\bm z^*=(1, \bm z^{\tiny \rm T})^{\tiny \rm T}$ (\citealp*{farewell1982use, kuk1992mixture, peng2000nonparametric}). As observed in the case of logistic regression, the logistic model works well when subjects are linearly separable  into the cure or susceptible groups with respect to covariates. However, problem arises when subjects cannot be separated using a linear boundary. Other options to model the incidence include assuming a probit link function ($\Phi^{-1}(\pi(\bm z))=\bm z^{*\tiny \rm T} \bm \beta$) or a complementary log-log link function ($\log[-\log(1-\pi(\bm z))]=\bm z^{*\tiny \rm T} \bm \beta$), where $\Phi$ is the cumulative distribution function of the standard normal distribution (\citealp{peng2003fitting, cai2012smcure, tong2012mixture}). However, these link functions do not offer non-linear separability and are  not sufficient to capture more complex effects of $\bm z$ on the incidence. Non-parametric strategies, e.g., the generalized Kaplan-Meier estimate at maximum uncensored failure time (\citealp{xu2014nonparametric}) to estimate the incidence part $\pi(\bm z)$ and the modified Beran-type estimator (\citealp{lopez2017nonparametric}) to estimate the latency part in a mixture cure model, are also considered in the literature. Again, applying these strategies to multiple covariates can be challenging. Therefore, there exists necessity to identify a group of classifiers which would be able to model the incidence part more effectively by allowing non-linear separating boundaries between the cured and non-cured subjects.  \\

To this end, the support vector machine (SVM) could be a reasonable choice. Introduced by \cite{cortes1995support}, the SVM is a machine learning algorithm that finds a hyperplane in multidimensional feature space that maximizes the separating space (margin) between two classes. The main advantage of the SVM is that it can separate nonlinear inseparable data by transforming it to a higher dimensional space using kernel trick. Consequently, this classifier is more robust and flexible than logit or probit link functions. Recently, \cite{li2020support} studied the effect of the covariates on the incidence $\pi(\bm z)$ by implementing the SVM. The new mixture model is seen to outperform existing cure rate models especially in the estimation of the incidence, and performs well for non-linearly separable classes and high dimensional covariates. However, \cite{li2020support} only considered data under non-informative right censoring mechanism. Motivated by this work, we propose to employ the SVM based  modeling to study the effects of covariates on the incidence part of the mixture cure rate model for survival data subject to interval-censoring.  \\

Unlike right-censored data, interval-censored data occur for a study where subjects are inspected at regular intervals, and not continuously \cite{Jodi22}. If a subject meets with the event of interest, the exact survival time is not observed and is only known that the event has occurred between two consecutive inspections. Interval-censored data marked by cure prospect are often observed in follow-up clinical studies (cancer biochemical recurrence or AIDS drug resistance) dealing with events having low fatality and patients monitored at regular intervals (\citealp{sun2007statistical, lindsey1998methods}). As in the case of right-censored data, some subjects may never encounter the event of interest, and are considered as cured. Mixture cure models with interval censored data are examined based on several estimation techniques for both semiparametric and non-parametric set-ups (\citealp{kim2008cure, ma2009cure, ma2010mixed, xiang2011mixture, aljawadi2012nonparametric}).\\

The rest of the article is arranged as follows. In Section \ref{sect2}, we discuss about the mixture cure rate model framework for interval-censored data and develop an estimation procedure based on the expectation maximization (EM) algorithm that employs the SVM to model the incidence part. In Section \ref{sect3}, a detailed simulation study is carried out to demonstrate the performance of our proposed model in terms of flexibility, accuracy and robustness. Comparisons of our model with the existing logistic regression based  mixture cure rate models are made in this section. The model performance is further examined and illustrated in Section \ref{sect4} through an interval censored data on smoking cessation. Finally, we end our discussion by some concluding remarks and possible future research directions in Section \ref{sect5}.  \\

\section{SVM based  mixture cure rate model with interval censoring}\label{sect2}

\subsection{\it Censoring scheme and modeling lifetimes}\label{2.1}

The  data we observe in situations with interval censoring are of the form $(L_i, R_i, \delta_i, \bm x_i, \bm z_i)$ for $i=1, \dots, n$, where $n$ denotes the sample size. For the $i$-th subject, $L_i$ denotes the last inspection time before the event and $R_i$ denotes the first subsequent inspection time just after the event. Note that $L_i < R_i$. The censoring indicator is denoted by $\delta_i=I(R_i<\infty)$, which takes the value 0 if $R_i= \infty$, meaning that the event is not observed for a subject before the last inspection time, and takes the value 1 if $R_i < \infty$, meaning that the event took place but its exact time is not known and is only known to belong to the interval $[L_i,R_i]$. Now, $\bm x_i$ and $\bm z_i$ are the respective $p$ dimensional and $q$ dimensional covariate vectors affecting the incidence and latency parts, respectively, of the mixture cure rate model. To demonstrate the effect of covariates on the latency part, we consider a proportional hazards structure to model the lifetime distribution of the susceptible or non-cured subjects. That is, for the susceptible subjects, we model the hazard function by
\begin{equation}\label{eq4}
h_u(t_i|\bm x_i)= h_0(t_i) \exp\left\{\bm x_i^{\tiny \rm T}\bm \gamma \right\},
\end{equation} 
where $\bm \gamma=(\gamma_1, \dots, \gamma_p)^{\tiny \rm T}$ is the $p$ dimensional regression parameter vector measuring the effects of $\bm x$ and  $h_0(\cdot)$ is the unspecified baseline hazard function. To facilitate our discussion, we assume the baseline hazard to be of the following form: $h_0(t_i)=\alpha t_i^{\alpha-1}$, where $\alpha>0$. One is of course free to use other forms for the baseline hazard. Therefore, we have
\begin{equation}\label{eq5}
h_u(t_i|\bm x_i)=  \alpha t_i^{\alpha-1} \exp\left\{\bm x_i^{\tiny \rm T}\bm \gamma \right\}.
\end{equation} 
Note that \eqref{eq5} turns out to be the hazard function of a Weibull distribution with shape parameter $\alpha$ and scale parameter $\{ e^{\bm x_i^{\tiny \rm T} \bm \gamma}\}^{-1/\alpha}$. Weibull distribution is a popular and flexible choice for modeling lifetimes or failure times in survival analysis. It is closed under proportional hazards family when the shape parameter remains constant, and it accommodates decreasing ($\alpha < 1$), constant ($\alpha=1$) and increasing ($\alpha > 1$) failure rates (\citealp{farewell1982use, tsodikov2003estimating, kleinbaum2010survival}). From (\ref{eq2}), the resulting survival function and density function of any subject in the study (irrespective of the cured status) are respectively given by
\begin{equation}\label{eq6}
S_p(t_i|\bm x_i, \bm z_i)= 1-\pi(\bm z_i) + \pi(\bm z_i) \exp \left\{-\left({t_i/m_i}\right)^{\alpha}\right\}
\end{equation} 
and 
\begin{equation}\label{eq7}
f_p(t_i|\bm x_i, \bm z_i)=  \pi(\bm z_i) \frac{\alpha t_i^{\alpha-1}}{m_i^\alpha}   \times \exp \left\{-\left({t_i/m_i}\right)^{\alpha}\right\},
\end{equation} 
where $m_i=\{ e^{\bm x_i^{\tiny \rm T} \bm \gamma}\}^{-1/\alpha}$. \\

\subsection{\it Form of the likelihood function}\label{2.2}

As missing observations are inherent to the problem set-up and model framework, we propose to employ the EM algorithm to estimate the unknown parameters  (\citealp{mclachlan2007algorithm, sy2000estimation, peng2000nonparametric, Bal16}). For implementing the EM algorithm, we need the form of the complete data likelihood function.  Let us define $\Delta_0=\{i: \delta_i=0\}$ and $\Delta_1=\{i: \delta_i=1\}$.  Missing observations that appear in this context are in terms of the cure indicator variable $J$, where $J$ is as defined in \eqref{eq1}. Note that $J_i$'s are all known to take the value 1 if $i \in \Delta_1$. However, if $i \in \Delta_0$, $J_i$ can either take 0 or 1, and is thus unknown or missing. Using these $J_i$'s as the missing data, we can define the complete data as $(L_i, R_i, \delta_i, J_i,\bm x_i, \bm z_i)$, for $i=1, \dots, n$, which contain both observed and missing data. Under the interval censoring mechanism, we can now express the complete data likelihood function and log-likelihood function as:
\begin{equation}\label{eq8}
L_c=\prod_{i \in \Delta_1} \left[\pi(\bm z_i) \left\{S_u(L_i|\bm x_i)-S_u(R_i|\bm x_i)\right\}\right]^{J_i} \times \prod_{i \in \Delta_0} (1-\pi(\bm z_i))^{1-J_i} \left\{ \pi(\bm z_i)S_u(L_i|\bm x_i)\right\}^{J_i}
\end{equation} 
and 
\begin{flalign}\label{eq9}
l_c&=\sum_{i \in \Delta_1} J_i\left[\log \pi(\bm z_i) + \log \left\{S_u(L_i|\bm x_i)-S_u(R_i|\bm x_i)\right\}\right]\nonumber \\  
&+\sum_{i \in \Delta_0} (1-J_i)\log(1-\pi(\bm z_i))+ J_i \left\{ \log \pi(\bm z_i)+ \log S_u(L_i|\bm x_i)\right\},
\end{flalign} 
where $S_u(t_i|\bm x_i)=\exp\left\{-\left({t_i/m_i}\right)^{\alpha}\right\}$ (\citealp{pal2017likelihood}). It can be further noted that 
\begin{flalign}\label{eq10}
l_c=l_{c1}+l_{c2},
\end{flalign} 
where
\begin{flalign}\label{eq11}
l_{c1}=\sum_{i =1}^n \left [J_i \log \pi(\bm z_i) + (1-J_i) \log (1-\pi(\bm z_i))\right ]
\end{flalign} 
is a function that depends on the incidence part only and 
\begin{flalign}\label{eq12}
l_{c2}= \sum_{i =1}^n \left [\delta_i \log \left\{S_u(L_i|\bm x_i)-S_u(R_i|\bm x_i)\right\} + (1-\delta_i) J_i  \log S_u(L_i|\bm x_i)\right ]
\end{flalign} 
is a function that depends on the latency part only; see \cite{Pal21New}.

\subsection{\it Modeling the incidence part with support vector machine}\label{2.3}

Let us assume that $J_i$ for $i \in \Delta_0$ are observed by some mechanism to assist our theory.  Support vector machine algorithm maximizes the linear or non-linear margin between the two closest points belonging to the opposite classification groups (cured and susceptible). That is, SVM solves the following optimization problem for $d_i; i=1, \dots, n$:
 
\begin{equation}\label{eq13}
\underset{d_1, \dots, d_n}{\max}\left[ -\frac{1}{2} \sum_{i=1}^n \sum_{j=1}^n d_i d_j (2J_i-1)(2J_j-1)\Phi_k(\bm z_i, \bm z_j) + \sum_{i=1}^n d_i \right]
\end{equation}
subject to the constraint $\sum_{i=1}^n (2J_i-1)d_i=0$ and $0\le d_i \le C$, for $i=1, \dots, n$, where $C$ is a parameter that trades off between the margin width and misclassification proportion. Smaller values of $C$ cause optimizer to look for a larger margin width allowing higher misclassification. $\Phi_k(.,.)$ is a symmetric positive semi definite kernel function, which we consider to be the radial basis function (RBF) given by $\Phi_k(\bm z_i, \bm z_j)=\exp\left\{-\frac{(\bm z_i-\bm z_j)^{\tiny \rm T}(\bm z_i-\bm z_j)}{\sigma^2}\right\}$. RBF is a popular choice of the kernel function owing to its robustness by implementing the idea that a linear classifier in higher dimension can be used as a non-linear classifier in lower dimension. The parameter $\sigma^2$ determines the kernel-width. Both hyper-parameters $C$ and $\sigma^2$ are to be tuned to obtain the highest classification accuracy using cross-validation methods (\citealp{chang2011libsvm}).  Grid search can be implemented to determine $C$ and $\sigma^2$. Low values of $\sigma^2$ result in overfitting and jagged separator, while high values of $\sigma^2$ result in more linear and smoother decision boundaries. Also, it is recommended to standardize the covariate vector $\boldsymbol z$.\\

The mapping $J_i$ to $2J_i-1$ converts  the respective 0 and 1s to -1 and +1s, which aids in formulation of the optimization problem under the SVM framework. 
Once $d_i$'s are obtained, we can derive a threshold $b$ as $b=\sum_{i=1}^n (2J_i-1) d_i \Phi_k(\bm z_i, \bm z_{j})-(2J_j-1)$, for some $d_j > 0$. For any new covariate vector $\bm z_{new}$, the optimal decision or classification rule is given by 
\begin{equation}\label{eq13.5}
\psi(\bm z_{new})=\sum_{i=1}^n d_i (2J_i-1)\Phi_k(\bm z_i, \bm z_{new}) - b.
\end{equation}
As suggested by \cite{li2020support}, the sequential minimal optimization method (SMO), introduced by \cite{Platt99a}, can be applied to solve (\ref{eq13}). As opposed to solving large quadratic optimization problems to train a SVM model, SMO solves a series of smallest possible quadratic problems. Thus, SMO is relatively time inexpensive algorithm. Any subject with covariate $\bm z_{new}$ is assigned to the susceptible group if $\psi(\bm z_{new})> 0$ and to the cured group if $\psi(\bm z_{new})< 0$.\\

In the given context, note that it is not enough to just classify subjects as being cured or susceptible. It is also of our interest to obtain the estimates of uncured probabilities $\pi(\bm z_i)$ or equivalently the cured probabilities $1-\pi(\bm z_i)$. For this purpose, we use the Platt scaling method to obtain an estimate of $\pi(\bm z_i)$ from the classification rule $\psi(.)$ (\citealp{platt1999probabilistic}). The estimate of $\pi(\bm z_i)$ by Platt scaling method is given by
\begin{equation}\label{eq13.6}
\hat{\pi}(\bm z_i)= \frac{1}{1+\exp\{A \psi(\bm z_i)+B\}},
\end{equation}
where $A$ and $B$ are obtained by maximizing the following function: 
\begin{equation}\label{eq13.7}
\sum_{i=1}^n (1- \zeta_i)[A \psi(\bm z_i) + B] - \log[1 + \exp\{A  \psi(\bm z_i) + B\}] .
\end{equation}
Here, 
\begin{equation}\label{eq13.8}
  \zeta_{i} =
    \begin{cases}
      \frac{n^{(1)}+1}{n^{(1)}+2}, & \text{ if } J_i=1\\
      \frac{1}{n^{(0)}+2}, & \text{ if } J_i=0,
    \end{cases}       
\end{equation}
and $n^{(1)}$ and $n^{(0)}$ represents the number of subjects in the susceptible and cured groups, respectively.  \\

We started our discussion on the SVM based  modeling of the incidence part above with the assumption that $J_i$s are observed and available for training purpose. However, in practice, the cure status $J_i$ is not known for $i \in \Delta_0$. Multiple imputation based approach can be applied here to obtain $\hat{\pi}(\bm z_i)$ with imputed values of $J_i$ for $i=1, \dots, n$. The steps are as follows: 
\begin{enumerate}
\item For a pre-defined integer $N^*$ and $n^*=1, 2, \dots, N^*$, generate $\{J^{(n^*)}_i:  i = 1, \dots, n\}$,  where $J^{(n^*)}_i$ is a Bernoulli random variable with success probability  $p_i^{(n^*)}$. The discussion on deriving $p_i^{(n^*)}$ is provided in Section \ref{2.4}.  
\item For the imputed data $\{J^{(n^*)}_i:  i =1, \dots, n\}$, obtain ${\hat{\pi}}^{(n^*)}(\bm z_i)$ as the estimate of  ${\pi}(\bm z_i)$ by the Platt scaling method given in (\ref{eq13.6}) for $n^*=1, 2, \dots, N^*$.
\item  $\hat{\pi}(\bm z_i)=\left(1/N^{*}\right) \sum_{n^*=1}^{N^*} {\hat{\pi}}^{(n^*)}(\bm z_i)$ is the final estimate of ${\pi}(\bm z_i)$. 
\end{enumerate}

\subsection{\it Development of the EM algorithm}\label{2.4}

The E-step in the EM algorithm involves finding the conditional expectation of the complete data log-likelihood function in (\ref{eq9}) given the current estimates (say, at the $(r+1)$-th iteration step) and the observed data, which is equivalent to finding the conditional expectation of $J_i$ given the observed data, $\pi(\bm z_i)$ and $(\alpha, \bm \gamma^{ \tiny \rm T})^{\tiny \rm T}$,  as
\begin{eqnarray}\label{eq14}
w_i^{(r+1)}&=&\delta_i+(1-\delta_i)\frac{\pi^{(r)}(\bm z_i)S_u^{(r)}(L_i|\bm x_i)}{1-\pi^{(r)}(\bm z_i)+\pi^{(r)}(\bm z_i)S_u^{(r)}(L_i|\bm x_i)},\ \ i=1,\ldots,n,
\end{eqnarray} 
where $S_u^{(r)}(L_i|\bm x_i) = \exp\left\{-\left({L_i/m_i^{(r)}}\right)^{\alpha^{(r)}}\right\}$ with $m_i^{(r)} = \{ e^{\bm x_i^{\tiny \rm T} \bm \gamma^{(r)}}\}^{-1/\alpha^{(r)}}$. Note that \eqref{eq14} implies that $w_i^{(r+1)} = 1$ for all $i \in \Delta_1$. We obtain the conditional expectation of $l_c$ by simply replacing $J_i$'s with $w_i^{(r+1)}$ in (\ref{eq9}). We denote the aforementioned conditional expectation by
\begin{equation}\label{eq13.9}
Q_c=Q_{c1}+Q_{c2},
\end{equation}
where 
\begin{equation}
    Q_{c1} = \sum_{i =1}^n \left [w_i^{(r+1)} \log \pi(\bm z_i) + (1-w_i^{(r+1)}) \log (1-\pi(\bm z_i))\right ]
\end{equation}
and 
\begin{equation}
    Q_{c2} = \sum_{i =1}^n \left [\delta_i \log \left\{S_u(L_i|\bm x_i)-S_u(R_i|\bm x_i)\right\} + (1-\delta_i) w_i^{(r+1)}  \log S_u(L_i|\bm x_i)\right ].
    \label{Q2}
\end{equation}

The M-step updates the parameters in $Q_{c1}$ and $Q_{c2}$. For $r=0, 1, \dots $, the procedure for the $(r+1)$-th iteration step of the EM algorithm is given below.  


\begin{enumerate}
\item  Carry out the multiple imputation technique, as described in Section \ref{2.3}, by considering $p_i^{(n^*)}=w_i^{(r+1)}$, for $n^* = 1, \dots, N^*$ and $i=1, \dots, n$. Obtain $\hat{\pi}^{(r+1)}(\bm z_i)=\left(1/N^{*}\right) \sum_{n^*=1}^{N^*} {\hat{\pi}}^{(n^*)}(\bm z_i)$ by applying the Platt scaling method with the classification rule $\psi(\cdot)$ defined in (\ref{eq13.5}). Recall that the classification rule is built based on the imputed data $\{J^{(n^*)}_i :  i =1, \dots, n\}$, where $J^{(n^*)}_i$ is a Bernoulli random variable with success probability  $p_i^{(n^*)}$. 

\item Obtain $(\alpha^{(r+1)}, \bm \gamma^{(r+1) \tiny \rm T})$ by maximizing the function $Q_{c2}$, as defined in \eqref{Q2}, with respect to $\alpha$ and $\boldsymbol\gamma$. That is, find

\begin{flalign}\label{eq15}
(\alpha^{(r+1)}, \bm \gamma^{(r+1) \tiny \rm T})^{\tiny \rm T}=\underset{\alpha, \bm \gamma}{\arg \max }\text{ }  Q_{c2}.
\end{flalign}

\item Check for the convergence as follows: 
\begin{equation*}
    || \boldsymbol\theta^{(r+1)} - \boldsymbol\theta^{(r)} ||^2_2 < \epsilon,
\end{equation*}
where $\boldsymbol\theta^{(k)} = (\overline{\pi^{(k)}}(\boldsymbol z),\alpha^{(k)},\bm \gamma^{(k) \tiny \rm T})^{\tiny\rm T}$, with
$\overline{\pi^{(k)}}(\boldsymbol z) = \frac{1}{n}\sum_{i=1}^n\pi^{(k)}(\boldsymbol z_i)$, $\epsilon > 0$ is some pre-determined and sufficiently small tolerance and $||\cdot||_2$ is the $L_2$-norm. If the above criterion is satisfied, then, stop the algorithm.  In this case, $\hat{\pi}^{(r+1)}(\bm z_i)$, for $i=1, \dots, n$, and $(\alpha^{(r+1)}, \bm \gamma^{(r+1) \tiny \rm T})^{\tiny \rm T}$ are the final pointwise estimates. On the other hand, if the above criterion is not met, continue to Step 4.

\item Update $w_i^{(r+1)}$ in \eqref{eq14} to 

\begin{flalign}\label{eq16}
w_i^{(r+2)}=\delta_i+(1-\delta_i)\frac{\hat{\pi}^{(r+1)}(\bm z_i)S^{(r+1)}_u(L_i|\bm x_i)}{1-\hat\pi^{(r+1)}(\bm z_i)+\hat \pi^{(r+1)}(\bm z_i)S_u^{(r+1)}(L_i|\bm x_i)},
\end{flalign} 
where $S_u^{(r+1)}(t_i|\bm x_i)=\exp \left\{-\left({t_i/m_i^{(r+1)}}\right)^{\alpha^{(r+1)}}\right\}$ and $m_i^{(r+1)} = \{ e^{\bm x_i^{\tiny \rm T} \bm \gamma^{(r+1)}}\}^{-1/\alpha^{(r+1)}}$.

\item Repeat steps 1-4 until convergence is achieved.
\end{enumerate}

\subsection{\it Calculating the standard errors}\label{2.5}
The standard errors are estimated by non-parametric bootstrapping. For $b'=1, \dots, B$, $b'$-th bootstrapped data set is obtained by resampling with replacement from the original data. The sample size of the $b'$-th bootstrapped data is the same as the original data. Then, we carry out steps 1-5 of the EM algorithm as detailed in Section \ref{2.4} to obtain the estimates of model parameters for each bootstrapped data. This gives us $B$ estimates for each model parameter. For each parameter, the standard deviation of these $B$ estimates provide an estimate of the  standard error of the parameter.

\subsection{\it Finding the initial values}\label{2.6}

To start the EM algorithm, we need to provide initial values of $\pi(\boldsymbol z_i)$, for $i=1,\ldots,n$, along with $\alpha$ and $\boldsymbol\gamma$. To come up with an initial guess of $\pi(\boldsymbol z_i)$, first, we can consider the censoring indicator $\delta_i, i=1,\ldots,n$, as the cure indicator (i.e., $\delta_i=0$ would imply $J_i=0$ and $\delta_i=1$ would imply $J_i=1$). Then, we can apply the SVM to come up with the classification rule, as given in \eqref{eq13.5}, and, finally, we apply the Platt scaling method, as given in \eqref{eq13.6}, to obtain $\pi(\boldsymbol z_i)$. To obtain an initial guess of the latency parameters $\alpha$ and $\boldsymbol\gamma$, we make use of the form of the survival function of the susceptible subjects, i.e., $S_u(t_i)=\exp \left\{-\left({t_i/m_i}\right)^{\alpha}\right\},$ where $m_i=\{ e^{\bm x_i^{\tiny \rm T} \bm \gamma}\}^{-1/\alpha}$. Note that this form implies that
\begin{equation*}
    \log\{-\log S_u(t_i)\} = \alpha \log t_i + \boldsymbol x_i^{\tiny \rm T}\boldsymbol\gamma, \ \ i=1,\ldots,n.
\end{equation*}
Hence, we can fit a linear regression model using $\log\{-\log S_u(t_i)\}$ as the response to obtain estimates of $\alpha$ and $\boldsymbol\gamma$, which can be used as the initial guesses. 
For this purpose, $S_u(t_i)$ can be the estimated using the non-parametric Kaplan-Meier estimates. Since the form of the data is interval censored, we can take $t_i=\frac{L_i+R_i}{2}$, if $R_i<\infty$, and take $t_i=L_i$, if $R_i=\infty$, for all $i=1,\ldots,n$. Note that this procedure may result in negative estimates of $\alpha$. As such, we can take the initial guess of $\alpha$ as 0.05 or 0.1 if the estimate of $\alpha$ turns out to be negative.

\section{Simulation study} \label{sect3}

\begin{figure}[ht!]
\centering
\includegraphics[scale=0.7]{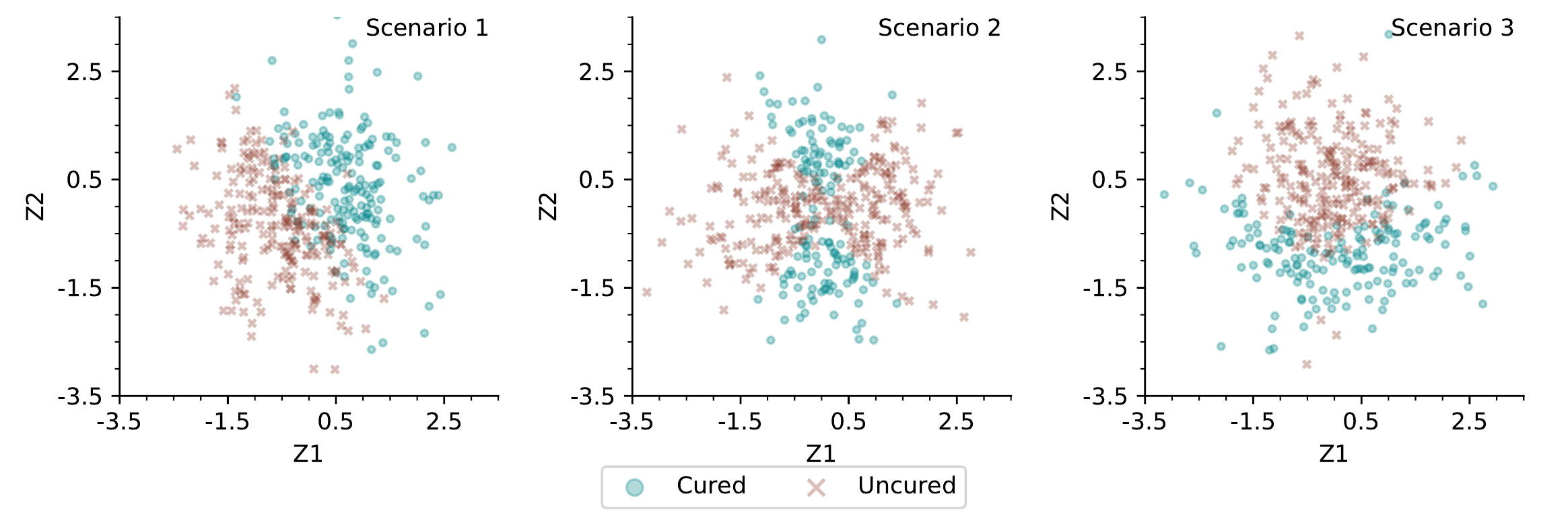}
\caption{Simulated cured and uncured observations for the three considered scenarios}
\label{figure:F1}
\end{figure}

\begin{figure}[ht!]
\centering
\includegraphics[scale=0.7]{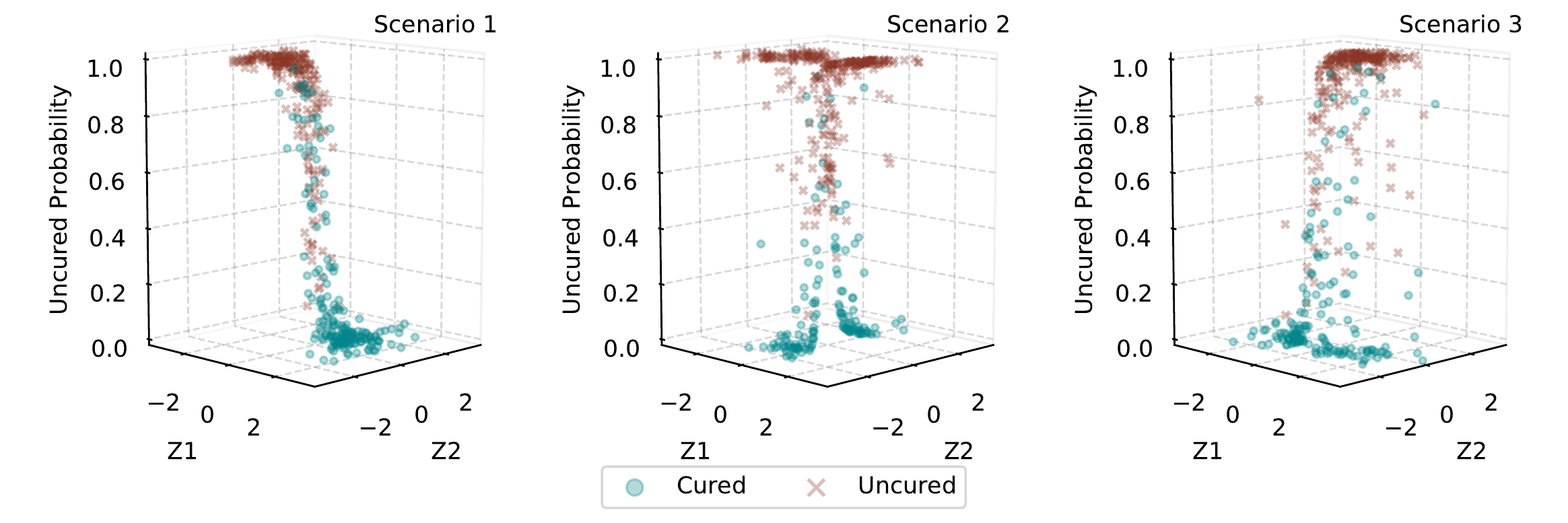}
\caption{Simulated uncured probabilities and their behavior with respect to the covariates for the three considered scenarios}
\label{figure:F2}
\end{figure}

In this section, we assess the performance of the proposed SVM based EM algorithm to estimate the model parameters of the mixture cure rate model for interval censored data. We generate two random values $x_1$ and $x_2$ independently from the standard normal distribution and assume $\bm x=\bm z$ with  $\bm x=(x_1, x_2)^{\tiny \rm T}$. We consider two different sample sizes: $n = 300$ and $n = 400$ and use the following links to generate uncured probabilities $\pi(\boldsymbol z)$:\\
\begin{eqnarray*}
\text{Scenario 1:} \ \ \pi(\boldsymbol z) &=& \frac{e^{0.3-5z_1-3z_2}}{1+e^{0.3-5z_1-3z_2}};\\
\text{Scenario 2:} \ \ \pi(\boldsymbol z) &=&  \frac{e^{0.3+10z_1^2-5z_2^2}}{1+e^{0.3+10z_1^2-5z_2^2}};\\
\text{Scenario 3:} \ \ \pi(\boldsymbol z) &=& \exp\{-\exp(0.3-4\cos z_1-5\sin z_2)\}.
\end{eqnarray*} 
Note that Scenario 1 represents the standard logistic regression model which captures a linear classification boundary. On the other hand, Scenarios 2 and 3 capture non-linear or more complex classification boundaries, as shown in Figure \ref{figure:F1}. Figure \ref{figure:F2} shows the plots of simulated uncured probabilities and how they vary with respect to the covariates $z_1$ and $z_2$. \\

We assume lifetimes of the susceptible subjects follow the proportional hazards structure with the hazard function
\begin{equation*}
h_u(t) = h_0(t)\exp(\gamma_1x_1+\gamma_2 x_2)
\end{equation*}
where $h_0(t) = \alpha t^{\alpha-1}$. As discussed before, the above hazard function implies that the susceptible lifetime follows a Weibull distribution with shape parameter $\alpha$ and scale parameter $\{ \exp(\gamma_1 x_1+\gamma_2 x_2)\}^{-\frac{1}{\alpha}}$.
We consider the true values of $(\alpha,\gamma_1,\gamma_2)$ as $(0.5,1,0.5)$. The censoring time is generated from a Uniform distribution in $(0,20)$. Under these settings, the cure probabilities range from $50\% - 65\%$, whereas the overall censoring proportions range from $60\% - 75\%$. To generate interval censored lifetime data $(L_i,R_i,\delta_i), i=1,2,\cdots,n$, we carry out the following steps:
\begin{enumerate}
\item [] Step 1: Generate a Uniform (0,1) random variable $U_i$ and a censoring time $C_i$;
\item [] Step 2: If $U_i\leq 1 - \pi(\boldsymbol z_i),$ set $L_i=C_i$, $R_i = \infty$, and $\delta_i = 0$;  
\item [] Step 3: If $U_i > 1 - \pi(\boldsymbol z_i),$ generate $T_i$ from a Weibull distribution with shape parameter $\alpha$ and scale parameter   $\{ \exp(\gamma_1 x_{1i}+\gamma_2 x_{2i})\}^{-\frac{1}{\alpha}}$; 
\item [] Step 4: 
\begin{enumerate}
\item [a.] If $\min\{T_i,C_i \} = C_i$,  set $L_i=C_i$, $R_i = \infty$, and $\delta_i = 0$;  
\item[b.]  If $\min\{T_i,C_i \} = T_i$,  set $\delta_i = 1$, and generate $L_{1i}$ from Uniform $(0.2,0.7)$ distribution and $L_{2i}$ from Uniform $(0,1)$ distribution. Next, create intervals $(0,L_{2i}], (L_{2i},L_{2i}+L_{1i}], \cdots, (L_{2i}+k\times L_{1i},\infty], k=1,2,\cdots,$ and select $(L_i,R_i)$ that satisfies $L_i<T_i\leq R_i$.
\end{enumerate}
\end{enumerate}

All simulations are done using the R statistical software (version 4.0.4) and all results are based on $M = 500$ Monte Carlo runs. To employ our proposed methodology, we consider  number of imputations in the multiple imputation technique to be 5, which is in line with \cite{li2020support}; see also \cite{wu2013cure}. In Table \ref{table:T1}, we report the bias and mean squared error (MSE) of the estimated uncured probability $\hat{\pi}(\boldsymbol z)$ and the susceptible survival probability $\hat{S_u}=\hat{S_u}(., .;\bm x)$. 
These are calculated as:
\begin{equation*}
\text{Bias}(\hat{\pi}(\boldsymbol z)) = \frac{1}{M}\sum_{k=1}^M \bigg [ \frac{1}{n}\sum_{i=1}^n \big\{ \widehat{\pi^{(k)}}(\boldsymbol z_i) - \pi^{(k)}(\boldsymbol z_i) \big\} \bigg];
\end{equation*}
\begin{equation*}
\text{Bias}(\hat{S_u}) = \frac{1}{M}\sum_{k=1}^M \bigg [ \frac{1}{n}\sum_{i=1}^n \big\{ \widehat{S_u^{(k)}}(L_i,R_i;\boldsymbol x_i) - S_u^{(k)}(L_i,R_i;\boldsymbol x_i) \big\} \bigg];
\end{equation*}
\begin{equation*}
\text{MSE}(\hat{\pi}(\boldsymbol z)) = \frac{1}{M}\sum_{k=1}^M \bigg [ \frac{1}{n}\sum_{i=1}^n \big\{ \widehat{\pi^{(k)}}(\boldsymbol z_i) - \pi^{(k)}(\boldsymbol z_i) \big\}^2 \bigg];
\end{equation*}
\begin{equation*}
\text{MSE}(\hat{S_u}) = \frac{1}{M}\sum_{k=1}^M \bigg [ \frac{1}{n}\sum_{i=1}^n \big\{ \widehat{S_u^{(k)}}(L_i,R_i;\boldsymbol x_i) - S_u^{(k)}(L_i,R_i;\boldsymbol x_i) \big\}^2 \bigg],
\end{equation*}
where $\pi^{(k)}(\boldsymbol z_i)$ and $S_u^{(k)}(L_i,R_i;\boldsymbol x_i)$ are the true uncured probability and susceptible survival probability, respectively, corresponding to the $i$-th subject and the $k$-th Monte Carlo run. Similarly, $\widehat{\pi^{(k)}}(\boldsymbol z_i)$ and $\widehat{S_u^{(k)}}(L_i,R_i;\boldsymbol x_i)$ are the estimated uncured probability and susceptible survival probability, respectively, corresponding to the $i$-th subject and the $k$-th Monte Carlo run. In the above expressions, note that $S_u^{(k)}(L_i,R_i;\boldsymbol x_i) = S_u^{(k)}(T_i;\boldsymbol x_i),$ where $T_i=\frac{L_i+R_i}{2}$ if $R_i<\infty$ and $T_i=L_i$ if $R_i=\infty$. $\widehat{S_u^{(k)}}(L_i,R_i;\boldsymbol x_i)$ is defined in a similar way. \\

\begin{table}[ht!]
\caption{Comparison of Bias and MSE of the uncured probability and susceptible survival probability}
\centering{
\resizebox{.95\textwidth}{!}{
\begin{tabular}{cccc|cc|cc|cc}
\hline
\multirow{3}{*}{$n$} & \multirow{3}{*}{Scenario} & \multicolumn{4}{c}{Uncured Probability} & \multicolumn{4}{c}{Susceptible Survival Probability}   \\ \cline{3-10}
& & \multicolumn{2}{c}{Bias} & \multicolumn{2}{c}{MSE} & \multicolumn{2}{c}{Bias} & \multicolumn{2}{c}{MSE} \\ \cline{3-10} 
& & SVM & LOGISTIC & SVM & LOGISTIC & SVM  & LOGISTIC & SVM & LOGISTIC \\ \hline
\multirow{3}{*}{400} & 1 & -0.126 & -0.002 & 0.083 & 0.002         & -0.062 & 0.001 & 0.021 & 0.001 \\
                     & 2 & -0.063 & 0.132 & 0.042 & 0.209          & -0.005 & 0.051  & 0.004 & 0.037    \\
                     & 3 & -0.020 & 0.089 & 0.019 & 0.080          & -0.006 & 0.013  & 0.002 & 0.005    \\ \hline
\multirow{3}{*}{300} 
                     & 1 & -0.126 & -0.001 & 0.088 & 0.002         & -0.063 & 0.002 & 0.022 & 0.001    \\
                     & 2 & -0.063 & 0.130  & 0.046 & 0.210         & -0.006  & 0.049 & 0.006 & 0.038    \\
                     & 3 & -0.023 & 0.087  & 0.022 & 0.080         & -0.006  & 0.013 & 0.003 & 0.006    \\ \hline
\end{tabular}
}
}
\label{table:T1}
\end{table}

\begin{figure}[ht!]
\centering
\includegraphics[scale=0.7]{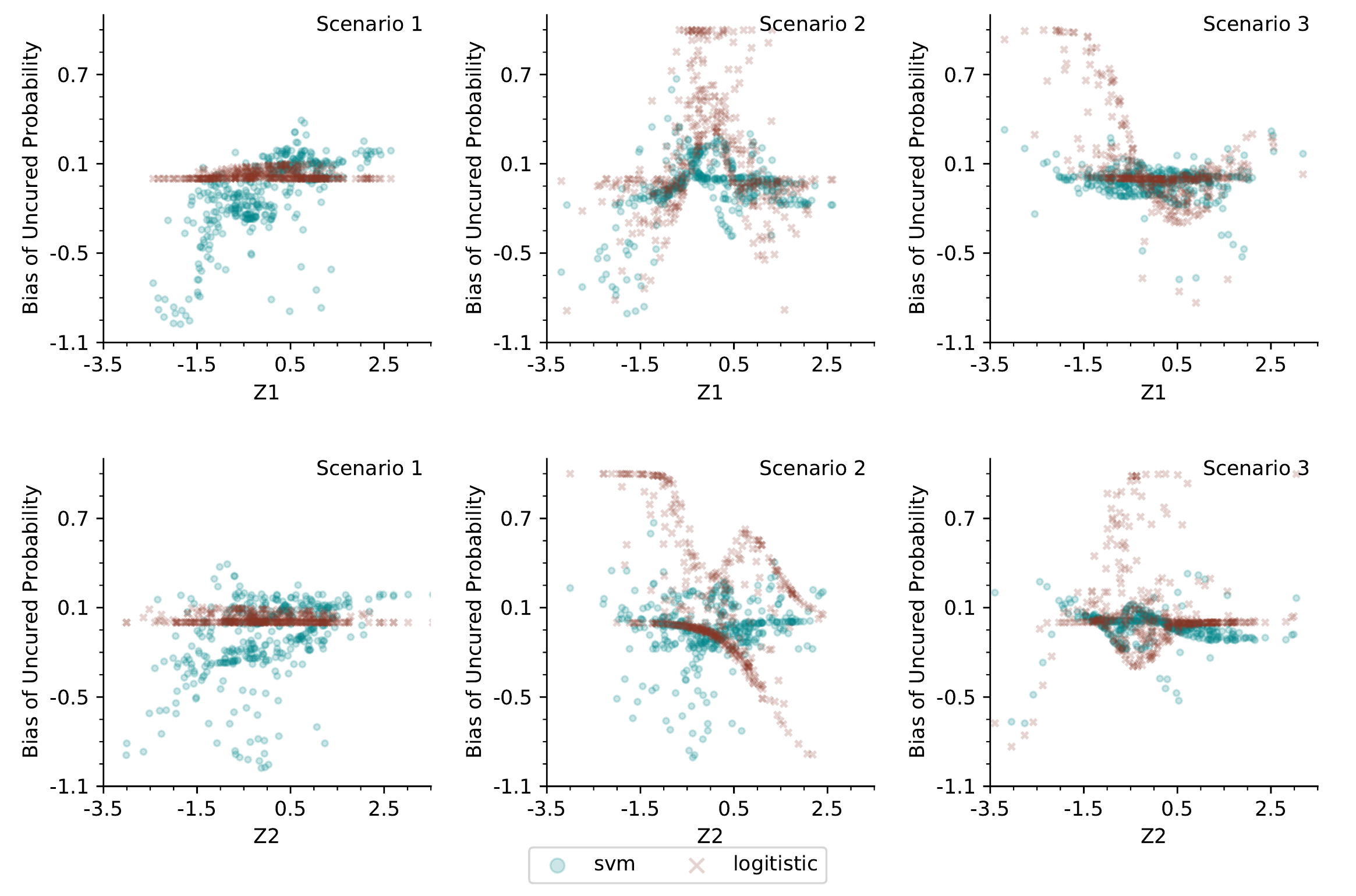}
\caption{Bias of the uncured probabilities with respect to each covariate for the three considered scenarios}
\label{figure:F3}
\end{figure}

From Table \ref{table:T1}, it is clear that the bias and MSE of the estimated uncured probability from the logistic based EM algorithm is smaller than that from the proposed SVM based EM algorithm when logistic regression is the correct model (Scenario 1). However, when the true model for the uncured probability is not the logistic regression in Scenarios 2 and 3, the proposed SVM based EM algorithm produces smaller bias and MSE in the estimated uncured probability. Figure \ref{figure:F3} presents the biases of the estimates of the individual uncured probabilities plotted against each covariate.\\ 

\begin{figure}[ht!]
\centering
\includegraphics[scale=0.65]{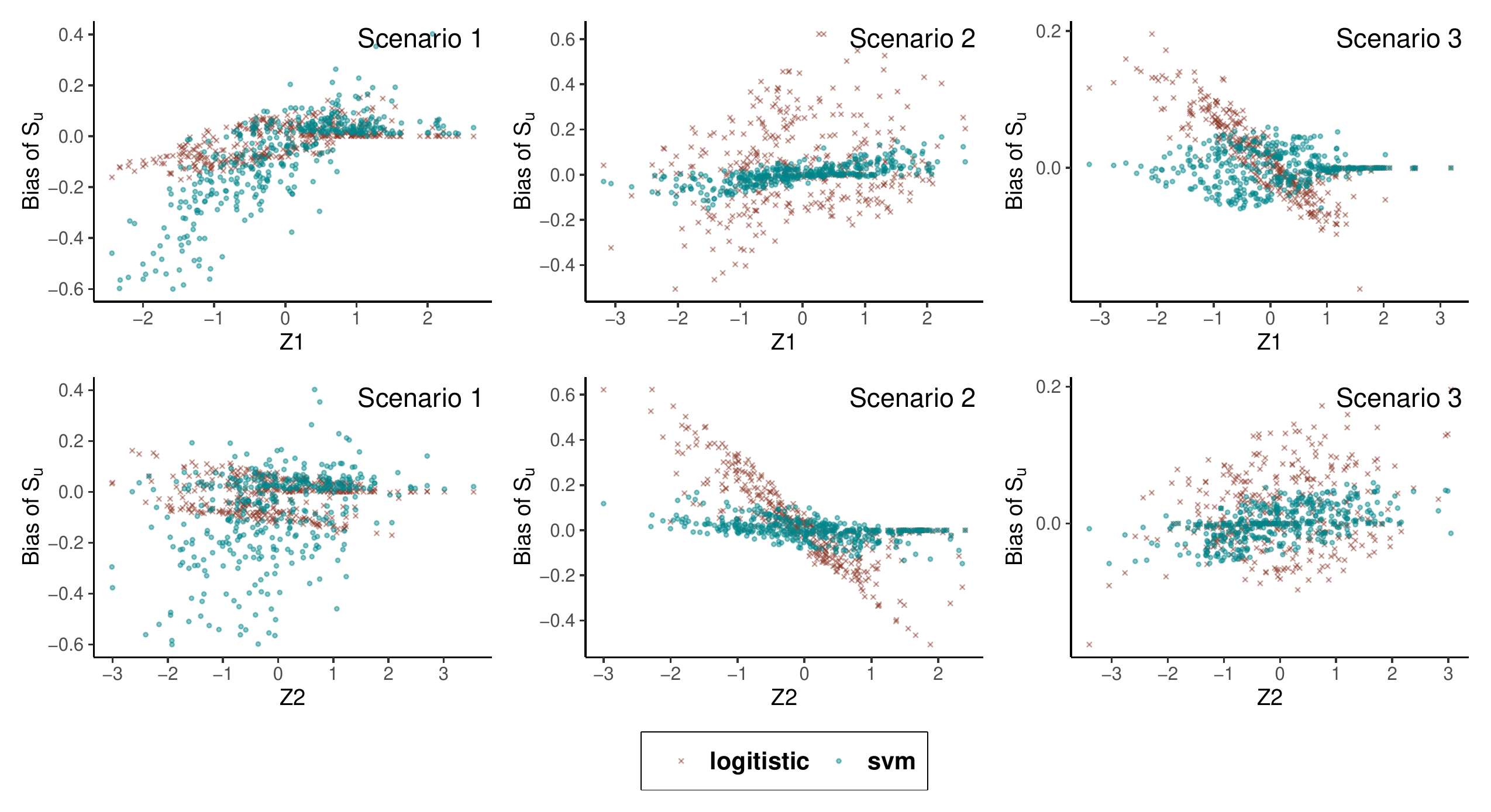}
\caption{Bias of the susceptible survival probabilities with respect to each covariate for the three considered scenarios}
\label{figure:F5}
\end{figure}

For the estimates of the susceptible survival probability, when the logistic regression model (Scenario 1) is the true model for the uncured probability, the logistic based EM algorithm produces smaller biases and MSEs compared to the SVM based EM algorithm. On the other hand, when the true model for the uncured probability is non-logistic (Scenarios 2 and 3), the SVM based  EM algorithm results in smaller biases and MSEs when compared to the logistic based  EM algorithm. Figure \ref{figure:F5} presents the biases of the estimates of the susceptible survival probabilities when plotted against each covariate. These findings clearly indicate that the SVM based EM algorithm is able to capture more complex and non-linear classification boundaries, where the standard logistic based EM algorithm produces relatively larger bias and MSE. \\

In Table \ref{table:T2}, we present the estimation results corresponding to the latency parameters. In particular, we compare bias, standard deviation (SD) and  MSE of the estimates of the latency parameters corresponding to the proposed SVM based  mixture cure rate model and the traditional logistic regression based mixture cure rate model. We can see that the bias, SD and MSE corresponding to the logistic regression based EM algorithm are  smaller when the logistic regression is the true model for the uncured probabilities (i.e., Scenario 1 is true). However, when the true model for the uncured probabilities is non-logistic (i.e., Scenarios 2 and 3 are the true models), the SVM based EM algorithm, in general, results in smaller bias, SD and MSE (note that in some cases, the estimates of parameters tend to have larger biases, SDs and MSEs in SVM method than in logistic method). With an increase in the sample size, the bias, SD and MSE tend to decrease further, which is what we would expect. \\

Summarizing the findings from both Table \ref{table:T1}
and Table \ref{table:T2}, we can conclude that the proposed SVM based  EM algorithm performs better than the standard logistic regression based EM algorithm, both in terms of the incidence part and the latency part of the mixture cure rate model, when the true classification boundary is non-liner and complex. This clearly demonstrates the ability of the proposed SVM based  model to handle complex non-linear classification boundaries. \\

\begin{figure}[ht!]
\centering
\includegraphics[scale=0.8]{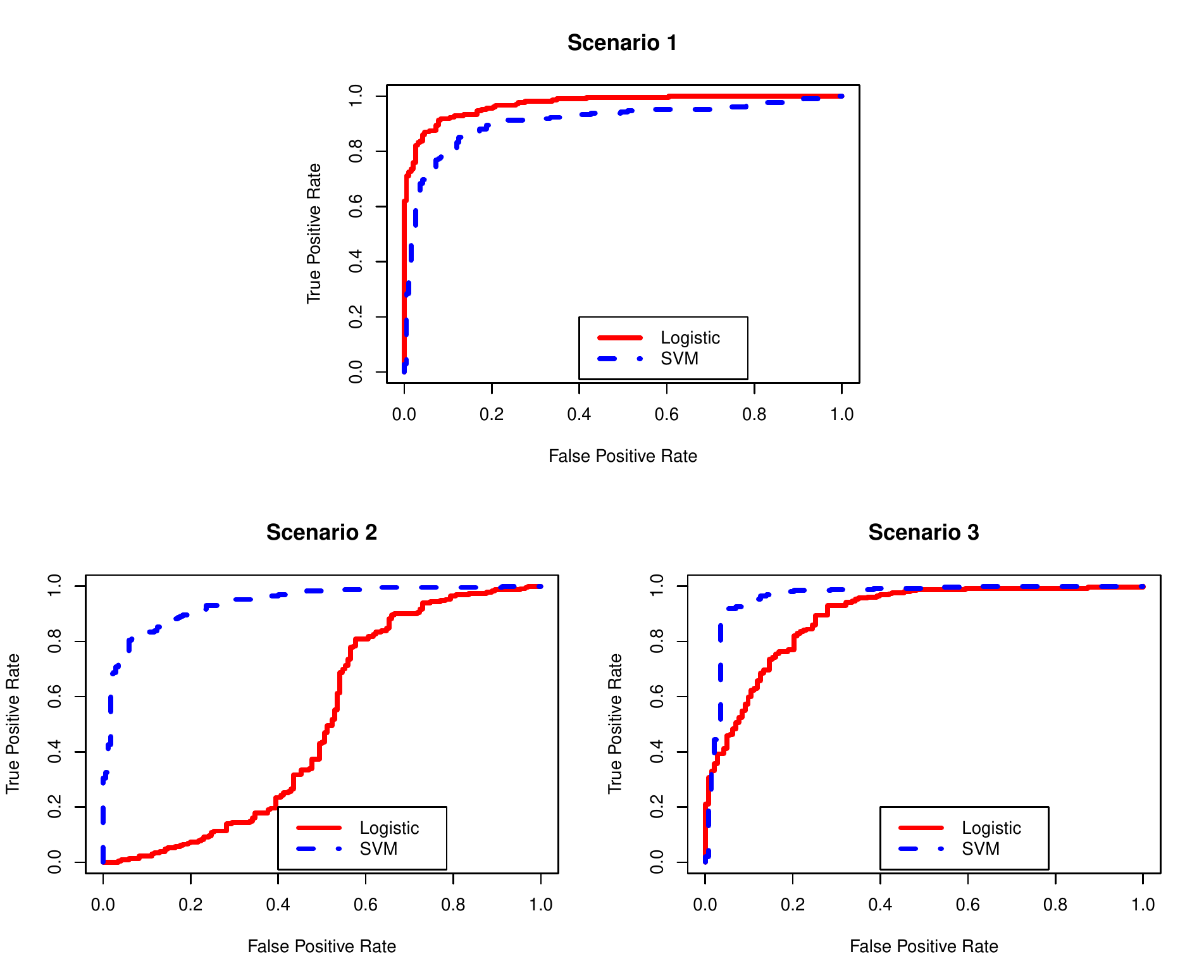}
\caption{ROC curves under different scenarios}
\label{figure:F7}
\end{figure}

Although, in practice, the cured status is unobserved for a real data, we do know which observations can be considered as cured when we simulate data. Using such information on the cured status for simulated data, we can easily compare the proposed SVM based  mixture model with the logistic regression based  mixture model using the receiver operating characteristic (ROC) curves and the area under the curves (AUCs) for different scenarios we have considered. Figure \ref{figure:F7} presents the ROC curves under different scenarios. The corresponding AUC values are presented in Table \ref{table:T3}. These results are based on 500 Monte Carlo runs with $n=400$ in each run. It is once again clear that under Scenarios 2 and 3 (i.e., when the classification boundaries are non-linear), the performance (or the accuracy) of the SVM based  model is better than the logistic regression based  model. Note, in particular, that the performance of the SVM based  model is significantly better under Scenario 2. However, under scenario 1 (i.e., when the classification boundary is linear), the logistic regression based  model performs slightly better than the SVM based  model. 
 
\begin{table}[ht!]
\caption{Estimation results corresponding to the latency parameters}
\centering{
\resizebox{.95\textwidth}{!}{
\begin{tabular}{ccc|cc|cc|cc}
\hline
\multirow{2}{*}{$n$} & \multirow{2}{*}{Scenario} & \multirow{2}{*}{Latency Parameter} & \multicolumn{2}{c}{Bias} & \multicolumn{2}{c}{SD} & \multicolumn{2}{c}{MSE} \\ \cline{4-9} 
                     &                    &                 & SVM    & LOGISTIC & SVM   & LOGISTIC & SVM   & LOGISTIC \\ \hline
\multirow{9}{*}{400} & \multirow{3}{*}{1} & $\alpha = 0.5$  & 0.103  & 0.008    & 0.052 & 0.050    & 0.014 & 0.003    \\
                     &                    & $\gamma_1 = 1.0$   & -0.498 & 0.010    & 0.139 & 0.123    & 0.270 & 0.018    \\
                     &                    & $\gamma_2 = 0.5$ & -0.269 & 0.004    & 0.109 & 0.105    & 0.086 & 0.012    \\ [1ex]
                     & \multirow{3}{*}{2} & $\alpha = 0.5$  & 0.074  & -0.117   & 0.056 & 0.038    & 0.008 & 0.016    \\
                     &                    & $\gamma_1 = 1.0$   & -0.099 & -0.111   & 0.102 & 0.109    & 0.018 & 0.026    \\
                     &                    & $\gamma_2 = 0.5$ & -0.012 & 0.740    & 0.167 & 0.132    & 0.022 & 0.574    \\ [1ex]
                     & \multirow{3}{*}{3} & $\alpha = 0.5$  & 0.047  & -0.010   & 0.049 & 0.045    & 0.005 & 0.002    \\
                     &                    & $\gamma_1 = 1.0$   & -0.037 & 0.257    & 0.141 & 0.120    & 0.018 & 0.082    \\
                     &                    & $\gamma_2 = 0.5$ & 0.085  & 0.079    & 0.121 & 0.106    & 0.017 & 0.018    \\ [1ex] \hline 
\multirow{9}{*}{300} & \multirow{3}{*}{1} & $\alpha = 0.5$  & 0.107  & 0.007    & 0.062 & 0.060    & 0.015 & 0.004    \\
                     &                    & $\gamma_1 = 1.0$   & -0.526 & 0.006    & 0.164 & 0.143    & 0.303 & 0.021    \\
                     &                    & $\gamma_2 = 0.5$ & -0.281 & -0.004   & 0.131 & 0.125    & 0.096 & 0.014    \\ [1ex]
                     & \multirow{3}{*}{2} & $\alpha = 0.5$  & 0.067  & -0.116   & 0.067 & 0.047    & 0.009 & 0.017    \\
                     &                    & $\gamma_1 = 1.0$   & -0.093 & -0.102   & 0.123 & 0.129    & 0.022 & 0.029    \\
                     &                    & $\gamma_2 = 0.5$ & 0.009  & 0.722    & 0.198 & 0.164    & 0.033 & 0.598    \\ [1ex]
                     & \multirow{3}{*}{3} & $\alpha = 0.5$  & 0.056  & -0.004   & 0.060 & 0.053    & 0.007 & 0.003    \\
                     &                    & $\gamma_1 = 1.0$   & -0.036 & 0.252    & 0.162 & 0.141    & 0.021 & 0.085    \\
                     &                    & $\gamma_2 = 0.5$ & 0.092  & 0.073    & 0.142 & 0.125    & 0.021 & 0.022    \\ \hline
\end{tabular}
}
}
\label{table:T2}
\end{table}

\begin{table}[ht!]
\center
\caption{AUC values under different scenarios}
\begin{tabular}{ccc}
\hline
Scenario & LOGISTIC & SVM \\
\hline
1 & 0.973 & 0.927 \\
2 & 0.502 & 0.948 \\
3 & 0.873 & 0.962 \\
\hline
\end{tabular}
\label{table:T3}
\end{table}


\subsection{Comparison with spline-based mixture cure model and using non-parametric baseline survival function}

\begin{figure}[htb!]
\caption{Figure 2: ROC curves for different mixture cure models (MCM) and sample sizes}
\includegraphics[scale=0.6]{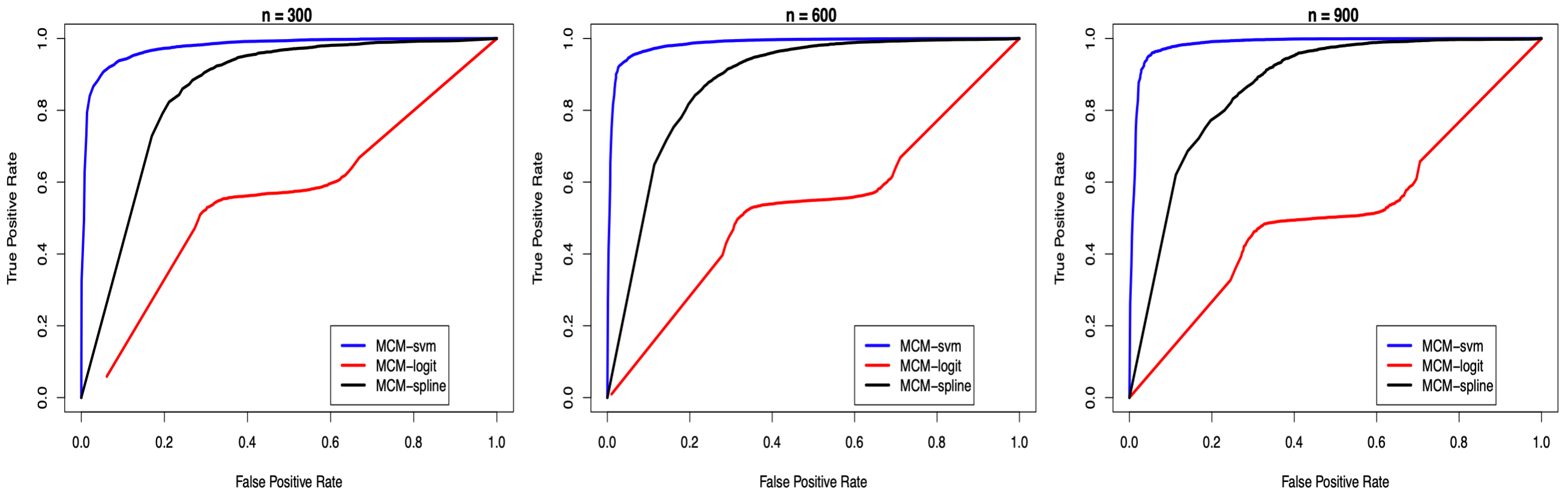}
\label{figure:FNEW}
\end{figure}

To demonstrate the superiority of our proposed model, we also compare our model with the spline regression-based mixture cure model which can also capture complex patterns in the data. We also relax the parametric assumption on the baseline hazard function and estimate the baseline survival function non-parametrically using the Turnbull type estimator. Considering scenario 3 and three different sample sizes ($n$=300, 600, 900), we present the results in Table \ref{table:New}. The corresponding ROC curves are presented in Figure \ref{figure:FNEW}. It is once again clear that our proposed SVM-based model performs better when compared to both spline-based and logistic regression-based models. 

\begin{table}[htb!]
\caption{Comparison of SVM-based model with spline-based and logistic regression-based models}
\begin{tabular*}{\textwidth}{@{\extracolsep{\fill}} l l l l l l l l l l}\\
\hline                                 
 &  \multicolumn{3}{c}{Bias}  &\multicolumn{3}{c}{MSE} & \multicolumn{3}{c}{AUC} \\  \cline{2-4} \cline{5-7} \cline{8-10}
n &SVM &Spline& Logit & SVM &Spline& Logit & SVM &Spline& Logit  \\
\hline
     300 & 0.0055 &0.0303& 0.0720 & 0.0174&0.0446& 0.0841 & 0.9748 & 0.8881 & 0.5932 \\
     600 & 0.0049 &0.0358& 0.0720 & 0.0124&0.0507& 0.0834 & 0.9838 & 0.8970 & 0.5723 \\
     900 & 0.0047 &0.0394& 0.0728 & 0.0097&0.0523& 0.0828 & 0.9851 & 0.8925 & 0.5470 \\
\hline
\end{tabular*}
\label{table:New}
\end{table}

\section{Illustrative example: smoking cessation data analysis}\label{sect4}

\begin{figure}[ht!]
\centering
\includegraphics[scale=0.6]{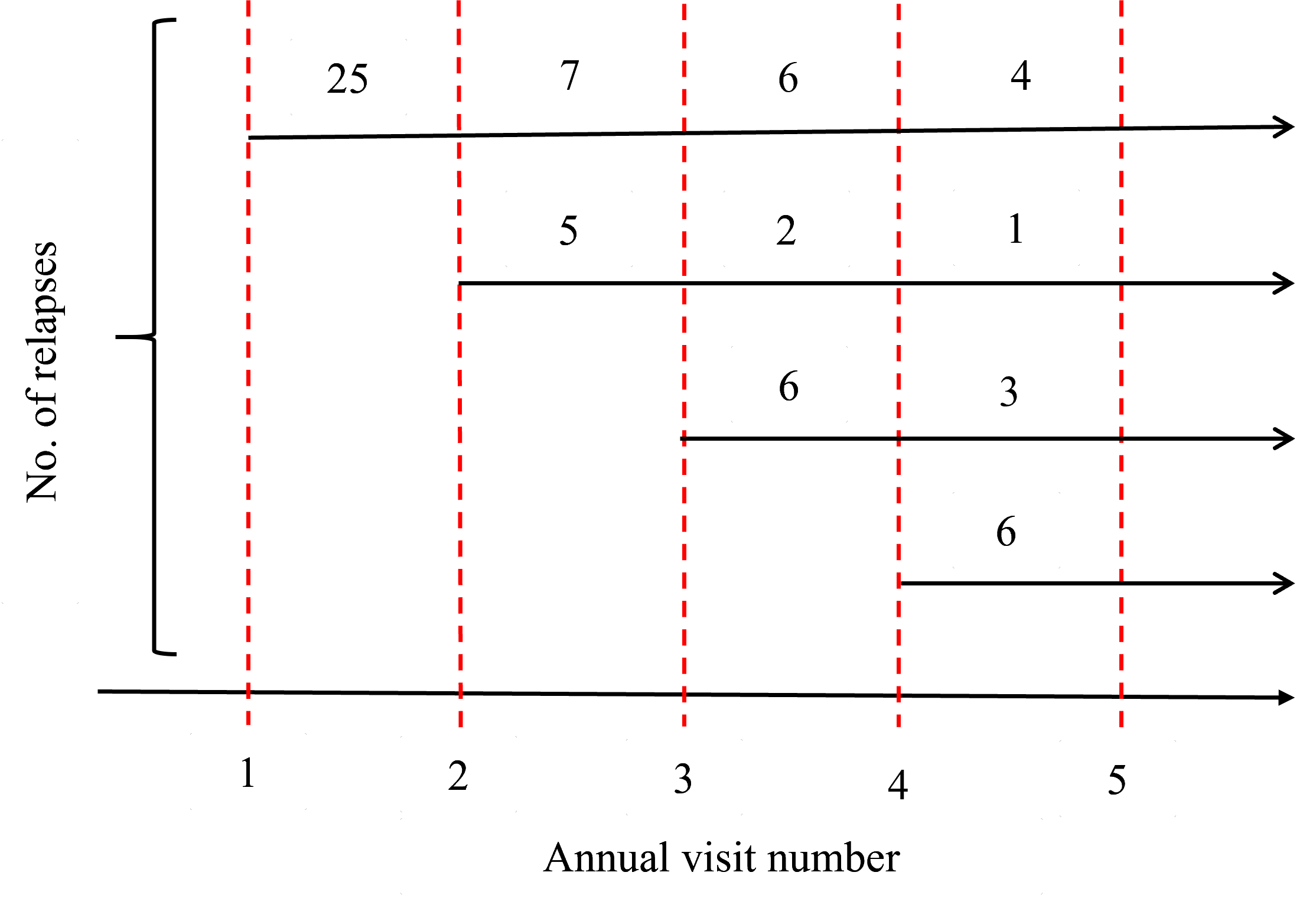}
\caption{Number of relapses in between every consecutive annual visits from study entry}
\label{figure:FS0}
\end{figure}

\begin{figure}[ht!]
\centering
\includegraphics[scale=0.6]{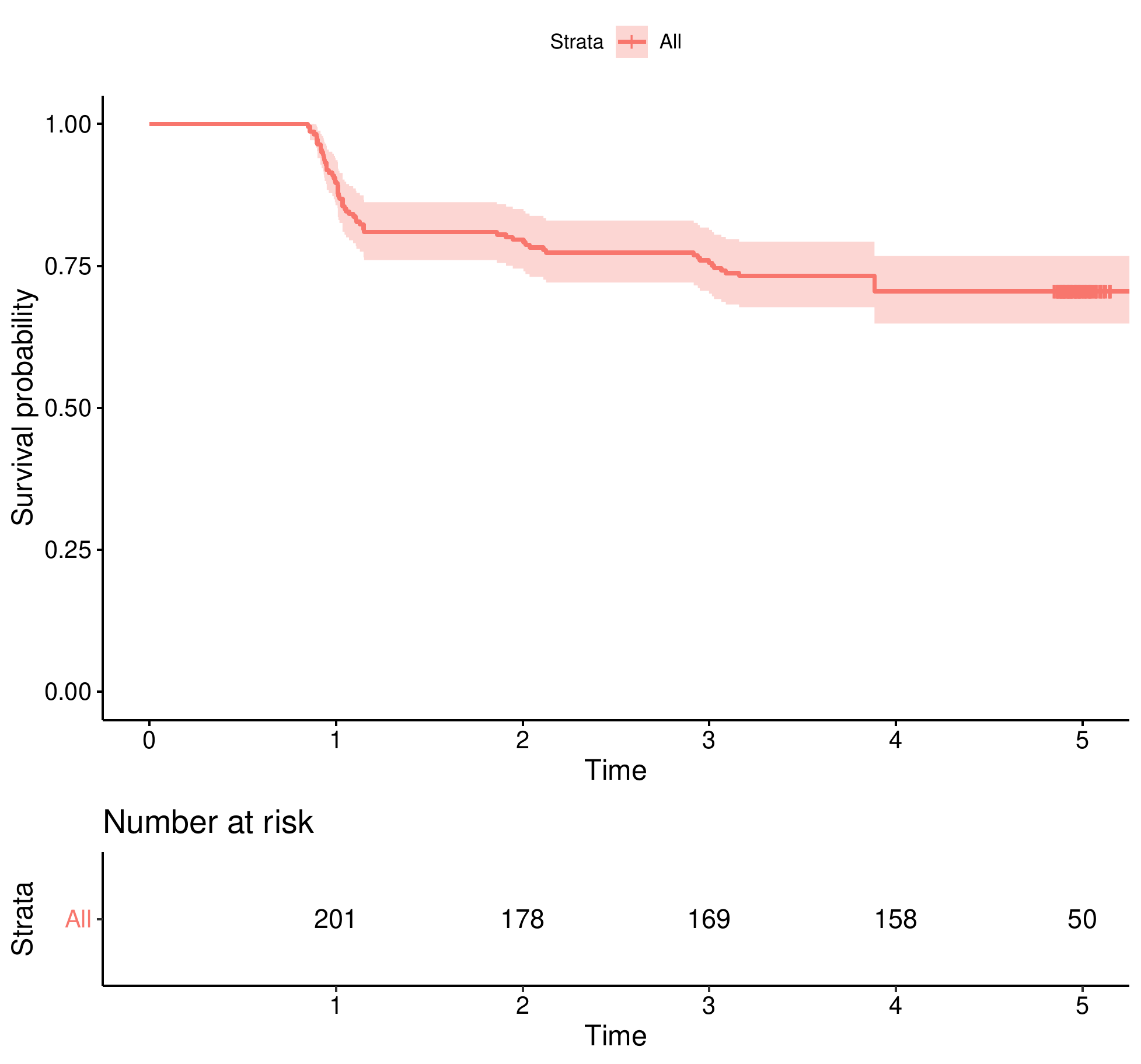}
\caption{Kaplan Meier curve for the smoking cessation data}
\label{figure:FS1}
\end{figure}

We further demonstrate our proposed methodology using a dataset on smoking cessation study (\citealp{murray1998effects,wiangnak2018gamma}). The study contains 223 subjects who had enrolled for the study during November 1986 to February 1989 (\citealp{banerjee2004parametric, kim2008cure}). Only those subjects who had tried to quit smoking at least once and who had identifiable Minnesota zip codes during the study period are considered in the analysis set. These subjects were all smokers at the time of enrollment, and were randomly assigned to two groups, namely, the smoking intervention (SI, treatment group) and the usual care (UC, control group). The subjects were monitored once every year for a period of 5 consecutive years. Information on whether they had relapsed or not (1:Yes and 0:No) are present in the data set. A relapse implies resumption of smoking and the event of interest for our illustration is the time to relapse. Obviously, the exact relapse time was unobserved since the relapse could have happened anytime in between two consecutive annual visits. Hence, the study falls under the scope of interval censored data analysis. Information on several additional variables are also available, e.g., gender (GEN, 1:Female and 0:Male), duration of smoking (DUR, time in years elapsed between commencement of smoking and entry to the study) and average number of cigarettes smoked per day (AVGCIG) before the  study period. These variables are treated as covariates since these factors supposedly can influence the relapse. Out of those who relapsed, most did so in the first year of their smoking cessation trial (see Figure \ref{figure:FS0}). In Figure \ref{figure:FS1}, we present the Kaplan-Meier curve. Clearly, we can see that the curve levels off to a significant non-zero proportion. This indicates that there could be a greater likelihood of the presence of cured fraction in the data. In Table \ref{table:RDA1}, we present few important descriptive statistics related to the study.\\

\begin{table}[ht!]
\caption{Distribution of proportion of relapse, average duration and average number of cigarettes smoked per year by gender and treatment group}
\centering{
\begin{tabular}{c|c|cc}
\hline
Treatment Group & Measure    &\multicolumn{2}{c}{Gender}  \\ 
\cline{3-4}
 & & Female & Male\\ \hline
& $n$ $(\%)$& 73 (32.735) & 96 (43.049)\\
SI & $\hat {p}_r$ ($95\%$ CI)  & 0.329 (0.221, 0.437) & 0.219 (0.136, 0.301)\\
 & Avg Dur (SD) & 29.506 (6.390) & 25.246 (9.667)\\
& Avg Cig (SD) & 30.343 (7.115) & 29.375 (12.552)\\
\hline
& $n$ $(\%)$& 14 (6.278) & 40 (17.937)\\
UC & $\hat {p}_r$ ($95\%$ CI)  & 0.357 (0.106, 0.608) & 0.375 (0.224, 0.525)\\
 & Avg Dur (SD) & 28.214 (8.833) & 22.714 (9.160)\\
& Avg Cig (SD) & 30.750 (7.502) & 26.875 (9.915)\\
\hline
\end{tabular}
}
\label{table:RDA1}\\
\footnotesize{SI: smoking intervention, UC: usual care, $n$: sample size, $\%$: percentage of the total, $\hat {p}_r$: proportion of relapse, CI: confidence interval, Avg Dur: average of DUR, Avg Cig: average of AVGCIG, SD: standard deviation}\\
\end{table}

\begin{figure}[hptb!]
\centering
\begin{tabular}{cc}
		\includegraphics[scale=0.6]{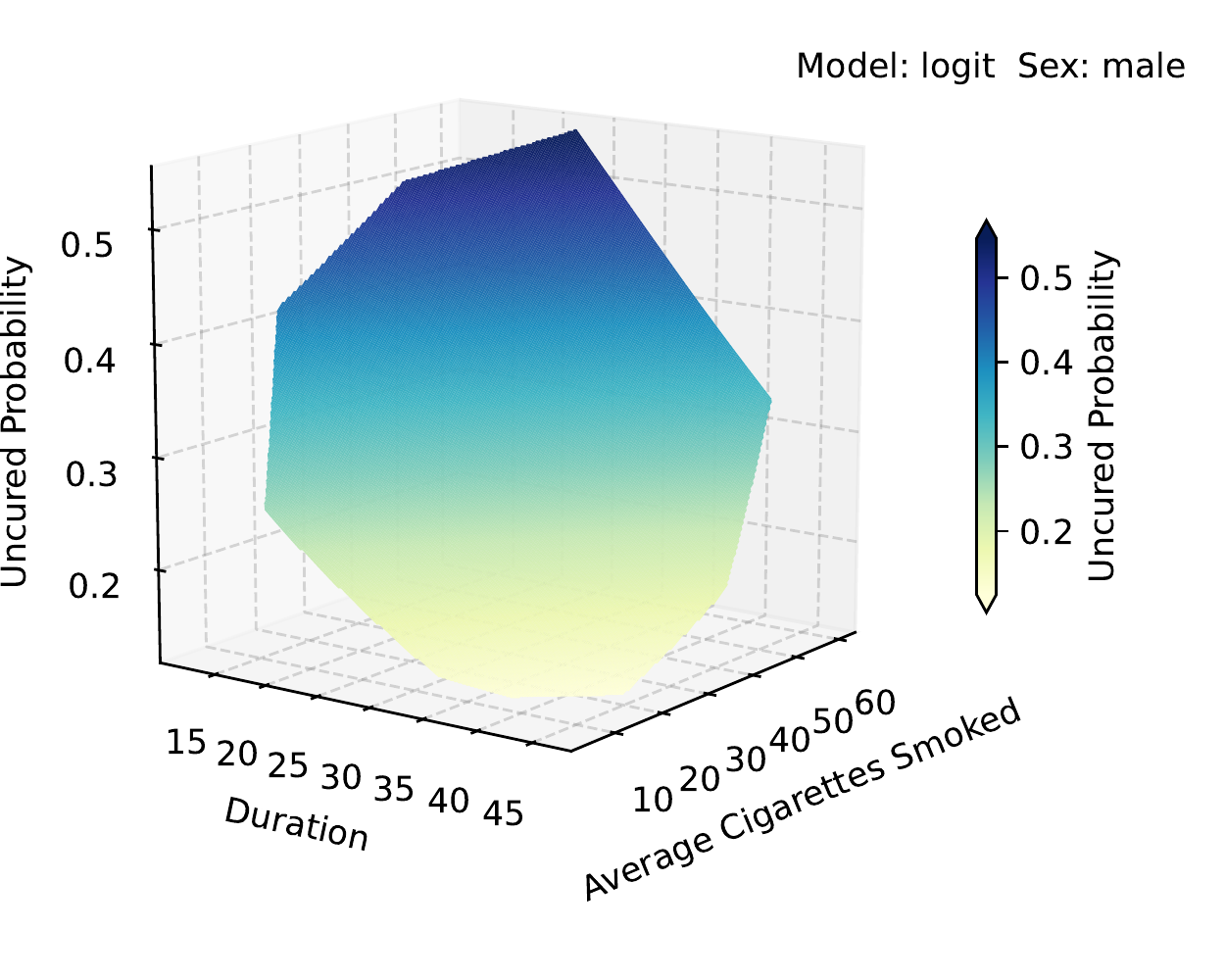}
		\includegraphics[scale=0.6]{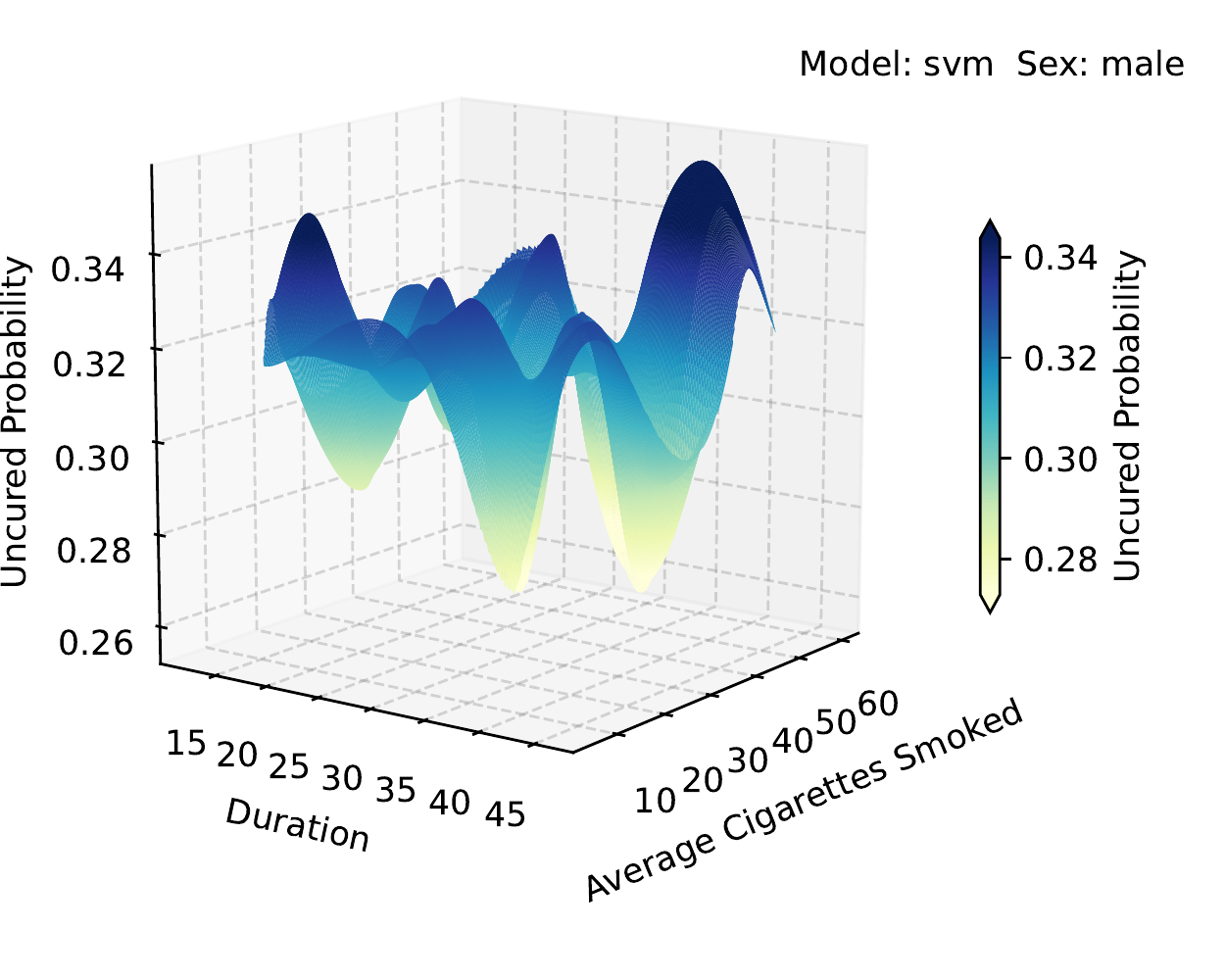}\\
		\includegraphics[scale=0.6]{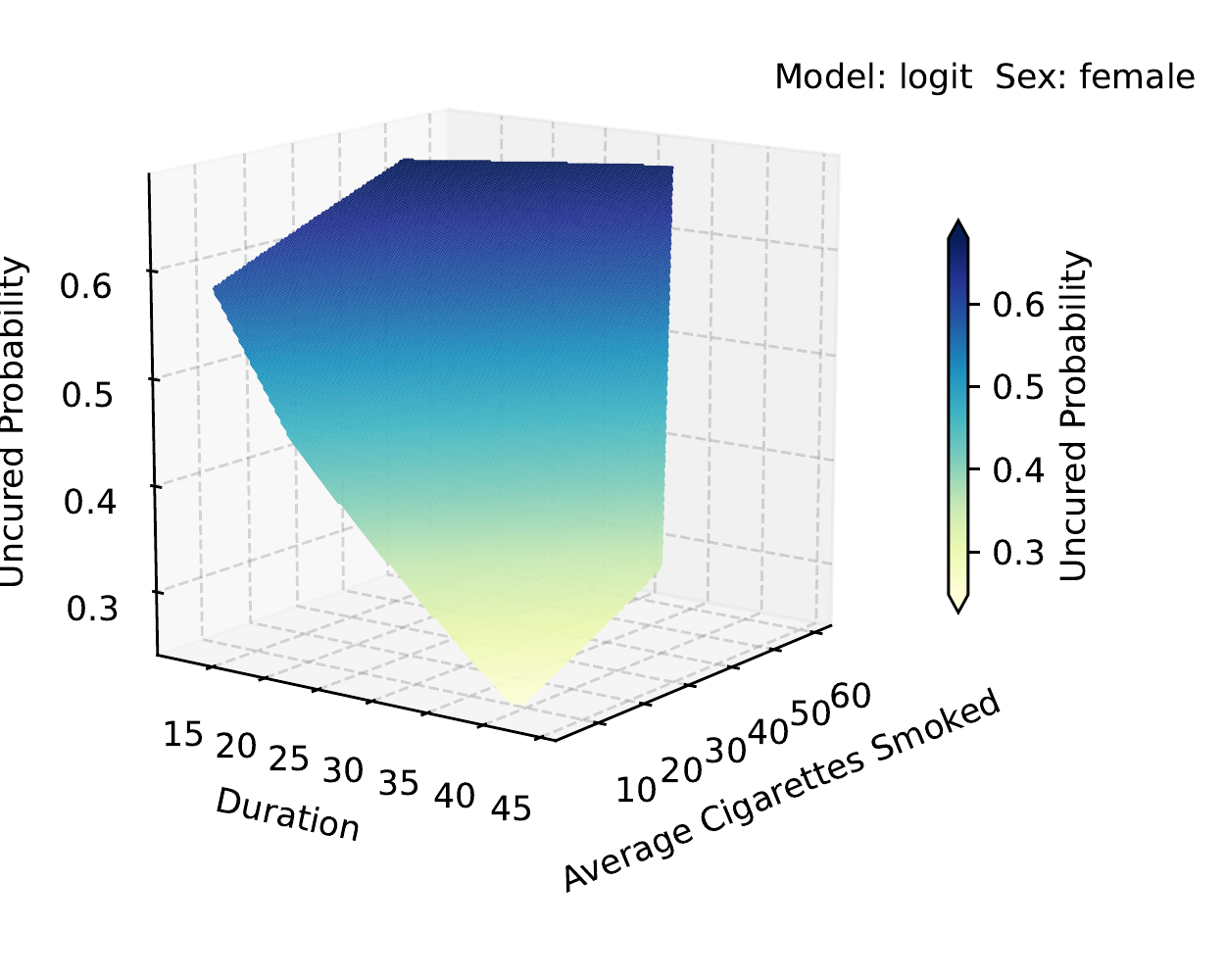}
		\includegraphics[scale=0.6]{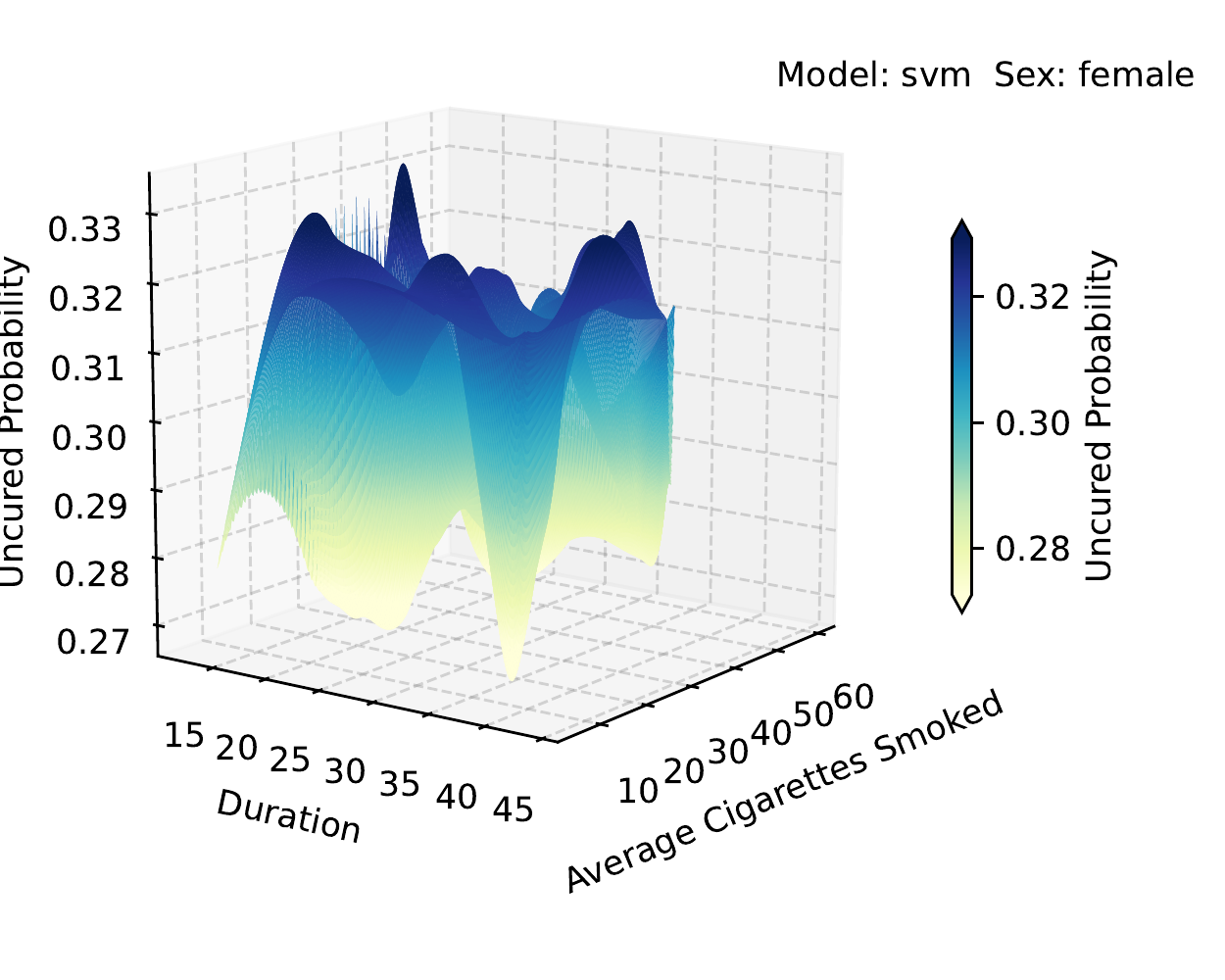}
\end{tabular}
  \caption{Estimates of uncured probabilities as a function of DUR and AVGCIG}
\label{figure:3D}
\end{figure}

In our application, we consider DUR ($x_1$), AVGCIG ($x_2$) and GEN ($x_3$) as covariates of interest. We fit the proposed SVM based  mixture cure rate model and, for comparison, we also fit the logistic regression based  mixture cure rate model. First, we draw inference on the incidence part of the model. In Figure \ref{figure:3D}, for each gender, we plot the estimates of the uncured probabilities against DUR and AVGCIG for both models. Clearly, under the proposed SVM based  model, the change in the estimates of the uncured probabilities is non-monotonic with respect to DUR and AVGCIG. This non-monotonic relationship is not captured by the logistic regression based model, owing to its rigid model assumption. \\

Table \ref{table:T4} presents the estimates of the latency parameters and their standard deviations for both SVM based and logistic regression based models. The effects of the covariates on the latency part are the same for both models. Clearly, at 1\% level of significance, only GEN turns out to be significant as far as the time to relapse of uncured patients is concerned. Since the estimate of $\gamma_3$ is negative, males tend to relapse faster than females. Also, since the estimate of $\gamma_1$ is positive, the hazard of smoking relapse increases with longer duration of smoking. However, such an effect is not significant. Moreover, since the estimate of $\gamma_2$ is negative, it implies that those who smoked less cigarettes tend to relapse faster. This effect is significant at 5\% level of significance only under the SVM based model. In the Appendix, we present two plots. Figure \ref{figure:FS2} presents the predicted survival probabilities of uncured subjects for fixed DUR and different values of AVGCIG. Figure \ref{figure:FS3} presents the predicted survival probabilities of uncured subjects for fixed AVGCIG and different values of DUR.



\begin{table}[ht!]
\center
\caption{Estimation results corresponding to the latency parameters for the smoking cessation data}
\begin{tabular}{lcc|cc|cc}
\hline
Parameter & \multicolumn{2}{c}{Estimates} & \multicolumn{2}{c}{SD} & \multicolumn{2}{c}{$p$-value} \\ \hline
& SVM & LOGISTIC & SVM & LOGISTIC & SVM  & LOGISTIC  \\ \hline
$\alpha$             & 1.013  & 0.968    & 0.129 & 0.093    & -- & -- \\
$\gamma_1$ (DUR)     & 0.176  & 0.210    & 0.143 & 0.206    & 0.218 & 0.307 \\
$\gamma_2$ (AVGCIG)  & -0.283 & -0.330   & 0.129 & 0.256    & 0.028 & 0.198 \\
$\gamma_3$ (GEN)     & -1.058 & -1.423   & 0.150 & 0.200    & 1.71$\times 10^{-12}$ & 1.32$\times 10^{-12}$ \\ \hline
\end{tabular}
\label{table:T4}
\end{table}

\section{Conclusion} \label{sect5}

The support vector machine has received a great amount of interest in the past two decades. It has been shown that the SVM performs well in a wide array of problems including face detection, text categorization and pedestrian detection. However, the use of the SVM in the context of cure rate models is new and not well explored. In this manuscript, we have proposed a new cure rate model that uses the SVM to model the incidence part and a proportional hazards structure to model the latency part for survival data subject to interval censoring. The new cure rate model inherits the properties of the SVM and can capture more complex classification boundaries. For the estimation purpose, we have proposed an EM algorithm where sequential minimal optimization together with Platt scaling method are employed to estimate the uncured probabilities. In this regard, due to the unavailability of some cured statuses, we make use of a multiple imputation based  approach to generate missing cured statuses. Due to the complexity of the proposed model and the estimation method, we approximate the standard errors of the estimated parameters using non-parametric bootstrapping. Through a simulation study, we have shown that when the true classification boundary is non-linear the proposed SVM based  model performs better than the standard logistic regression based  model. This is true with respect to both incidence and latency parts of the model. As future research, it is of great interest for us to extend the proposed model to accommodate a competing risks scenario (\citealp{balakrishnan2015algorithm, davies2020stochastic}). It is also of interest to explore other machine learning algorithms (e.g., neural network or tree-based approaches) to study more complicated cure rate models such as those that look at the elimination of risk factors (\citealp{Pal16,pal2017likelihood,Pal17a,Pal18b,majakwara2019some}) and those that belong to a transformation family of cure models \cite{Wang22}. We are currently looking at some of these problems and we hope to report the findings in our upcoming manuscripts.


\section*{Conflict of interest}

The authors declare that there is no conflict of interests related to the publication of this manuscript.

\bibliographystyle{apacite} 
\bibliography{crsvm}

\pagebreak

\appendix
\renewcommand\thefigure{\thesection A.\arabic{figure}} 
\section*{Appendix}

\setcounter{figure}{0}
\begin{figure}[ht!]
\centering
\includegraphics[scale=0.75]{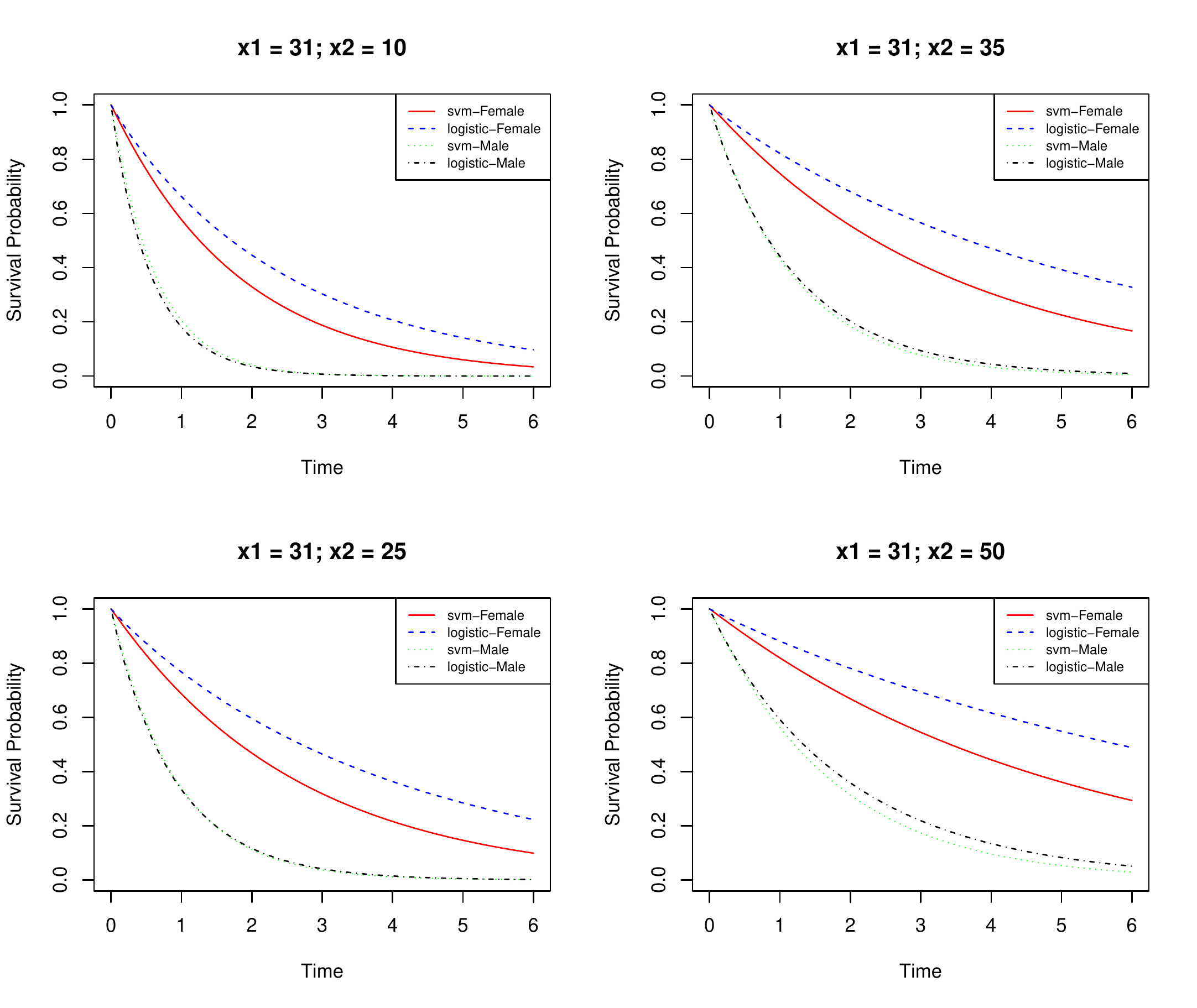}
\caption{Predicted survival probability of the susceptible for fixed duration as smoker $(x_1)$ and different values of average cigarettes smoked per day $(x_2)$}
\label{figure:FS2}
\end{figure}

\begin{figure}[ht!]
\centering
\includegraphics[scale=0.75]{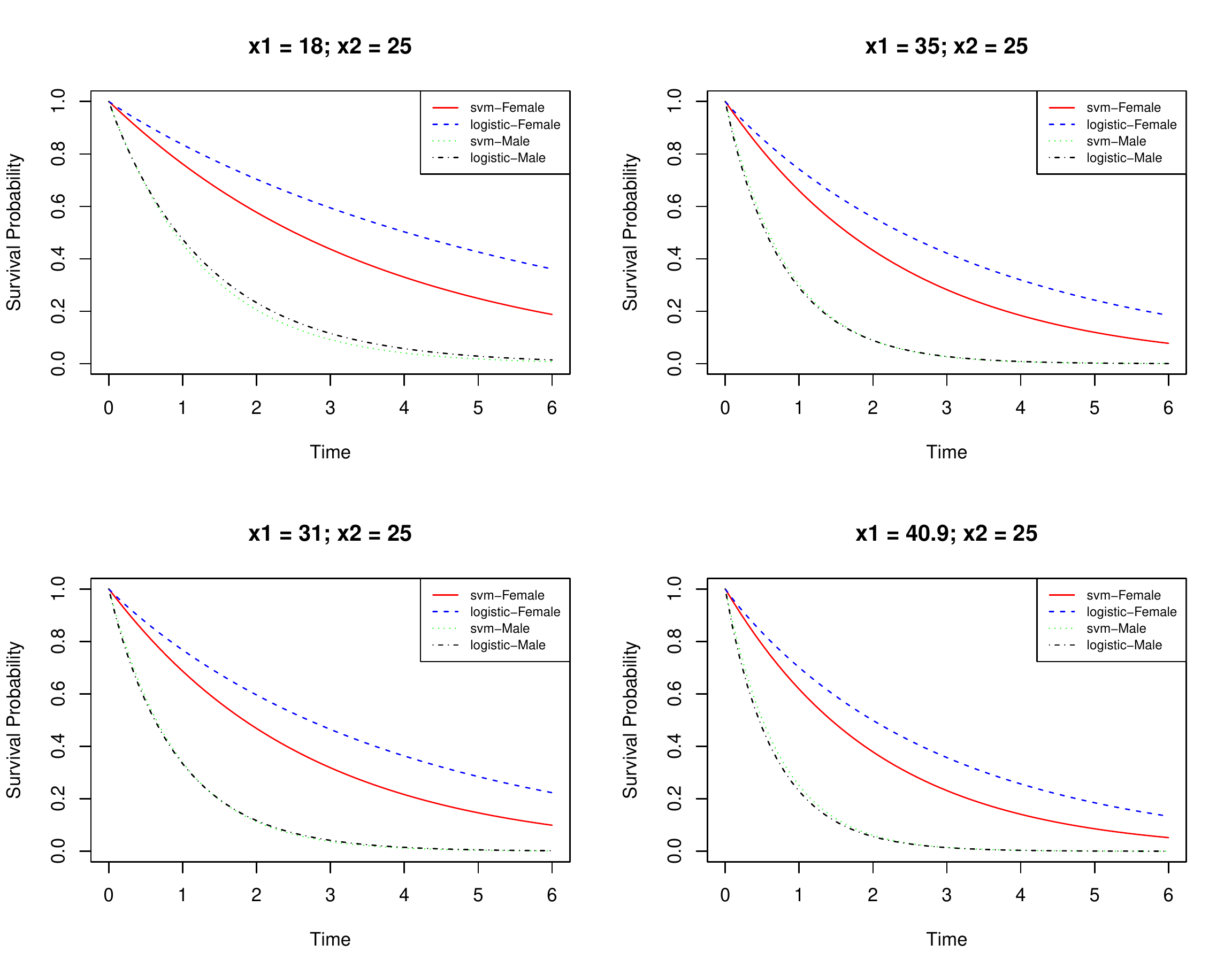}
\caption{Predicted survival probability of the susceptible for fixed average cigarettes smoked per day $(x_2)$ and different values of duration as smoker $(x_1)$}
\label{figure:FS3}
\end{figure}

\end{document}